\documentclass[12pt]{article}
\usepackage{amsmath}
\usepackage{graphicx}
\usepackage{axodraw}
\usepackage{subfigure}
\begin{document}
\baselineskip 0.6cm

\def\simgt{\mathrel{\lower2.5pt\vbox{\lineskip=0pt\baselineskip=0pt
           \hbox{$>$}\hbox{$\sim$}}}}
\def\simlt{\mathrel{\lower2.5pt\vbox{\lineskip=0pt\baselineskip=0pt
           \hbox{$<$}\hbox{$\sim$}}}}
\def\simprop{\mathrel{\lower5.0pt\vbox{\lineskip=1.0pt\baselineskip=0pt
             \hbox{$\propto$}\hbox{$\sim$}}}}
\newcommand{\vev}[1]{\langle #1 \rangle}

\begin{titlepage}

\begin{flushright}
UCB-PTH-12/03 \\
\end{flushright}

\vskip 1.5cm

\begin{center}

{\Large \bf 
Supersymmetry with Light Stops
}

\vskip 0.8cm

{\large Grant Larsen, Yasunori Nomura, and Hannes L.L. Roberts}

\vskip 0.4cm

{\it Berkeley Center for Theoretical Physics, Department of Physics, \\
     and Theoretical Physics Group, Lawrence Berkeley National Laboratory, \\
     University of California, Berkeley, CA 94720, USA} \\

\abstract{Recent LHC data, together with the electroweak naturalness 
 argument, suggest that the top squarks may be significantly lighter 
 than the other sfermions.  We present supersymmetric models in which 
 such a split spectrum is obtained through ``geometries'': being 
 ``close to'' electroweak symmetry breaking implies being ``away from'' 
 supersymmetry breaking, and vice versa.  In particular, we present 
 models in 5D warped spacetime, in which supersymmetry breaking and 
 Higgs fields are located on the ultraviolet and infrared branes, 
 respectively, and the top multiplets are localized to the infrared 
 brane.  The hierarchy of the Yukawa matrices can be obtained while 
 keeping near flavor degeneracy between the first two generation sfermions, 
 avoiding stringent constraints from flavor and $CP$ violation.  Through 
 the AdS/CFT correspondence, the models can be interpreted as purely 
 4D theories in which the top and Higgs multiplets are composites of 
 some strongly interacting sector exhibiting nontrivial dynamics at 
 a low energy.  Because of the compositeness of the Higgs and top 
 multiplets, Landau pole constraints for the Higgs and top couplings 
 apply only up to the dynamical scale, allowing for a relatively heavy 
 Higgs boson, including $m_h = 125~{\rm GeV}$ as suggested by the 
 recent LHC data.  We analyze electroweak symmetry breaking for 
 a well-motivated subset of these models, and find that fine-tuning 
 in electroweak symmetry breaking is indeed ameliorated.  We also 
 discuss a flat space realization of the scenario in which supersymmetry 
 is broken by boundary conditions, with the top multiplets localized 
 to a brane while other matter multiplets delocalized in the bulk.}

\end{center}
\end{titlepage}

\section{Introduction}
\label{sec:intro}

One of the strongest motivations for weak scale supersymmetry is the 
possibility of making electroweak symmetry breaking ``natural,'' i.e.\ 
a generic parameter region of the theory reproduces observed electroweak 
phenomena.  With the Higgs potential $V(h) = m^2 h^\dagger h + \lambda 
(h^\dagger h)^2/4$, the minimization of the potential leads to $v \equiv 
\vev{h} = \sqrt{-2m^2/\lambda}$ and
\begin{equation}
  \frac{m_h^2}{2} = -m^2,
\label{eq:Higgs-mass}
\end{equation}
where $m_h$ is the physical Higgs boson mass.  In the Standard Model 
(SM) a generic size of $|m^2|$ is expected to be at a scale where the 
theory breaks down, while in supersymmetric models
\begin{equation}
  m^2 = |\mu|^2 + \tilde{m}_h^2,
\label{eq:susy}
\end{equation}
where $\mu$ and $\tilde{m}_h^2$ are the supersymmetric and 
supersymmetry-breaking masses for the Higgs field.  Therefore, 
as long as these parameters are both of order the weak scale, 
the theory can naturally accommodate electroweak symmetry breaking.

Improved experimental constraints over the past decades, however, have 
cast doubt on this simple picture.  In softly broken supersymmetric 
theories, supersymmetry-breaking masses are affected by each other 
through renormalization group evolution; in particular, $\tilde{m}_h$ 
receives a contribution
\begin{equation}
  \delta \tilde{m}_h^2 \simeq -\frac{3m_t^2}{4\pi^2 v^2}\, 
    m_{\tilde{t}}^2 \ln\frac{M_{\rm mess}}{m_{\tilde{t}}},
\label{eq:soft}
\end{equation}
where $m_t$ and $m_{\tilde{t}}$ are the top quark and squark masses, 
and $M_{\rm mess}$ the scale at which supersymmetry breaking masses 
are generated.  (Here, we have ignored possible scalar trilinear 
interactions and set the left- and right-handed squark masses equal, 
for simplicity.)  Requiring no more fine-tuning than $\Delta$, 
Eqs.~(\ref{eq:Higgs-mass}) and (\ref{eq:susy}) lead to
\begin{equation}
  m_{\tilde{t}} \simlt 420~{\rm GeV}\, 
    \biggl( \frac{m_h}{125~{\rm GeV}} \biggr) 
    \biggl( \frac{20\%}{\Delta^{-1}} \biggr)^{1/2} 
    \biggl( \frac{3}{\ln\frac{M_{\rm mess}}{m_{\tilde{t}}}} \biggr)^{1/2}.
\label{eq:stop}
\end{equation}
On the other hand, recent observations at the LHC indicate:
\begin{itemize}
\item
Generic lower bounds on the first two generation squark masses are 
about $1~{\rm TeV}$~\cite{Aad:2011ib}.
\item
There are hints of the SM-like Higgs boson with $m_h \simeq 125~{\rm 
GeV}$~\cite{LHC:2011}.
\end{itemize}
Therefore, if the hints for the Higgs boson mass are true, then it 
strongly suggests that the squark masses have a nontrivial flavor 
structure, i.e.\ top squarks (stops) are light.%
\footnote{One way of avoiding this conclusion is to invoke a 
 significant mixing of the Higgs field with another scalar field; 
 see~\cite{Hall:2011aa}.  In general, mixing of the SM-like 
 Higgs field with another field can weaken the naive constraint, 
 Eq.~(\ref{eq:stop}), obtained in the decoupling regime (at the cost 
 of moderate cancellation in a scalar mass-squared eigenvalue). 
 Another possibility is to have a relatively compressed superparticle 
 spectrum, in particular a small mass splitting between the squarks 
 and the lightest neutralino, in which case the lower bound on the 
 (light generation) squark masses becomes weaker.}

The above observation has significant implications on an underlying 
model of supersymmetry breaking.  This is especially because many 
existing models, including minimal supergravity, gauge mediation, 
and anomaly mediation, invoke flavor universality to avoid stringent 
constraints from the absence of large flavor violating processes. 
On the other hand, it has been realized that naturalness itself 
allows sfermions other than the stops (and the left-handed 
sbottom) to be significantly heavier than the value suggested by 
Eq.~(\ref{eq:stop})~\cite{Dine:1990jd,Dimopoulos:1995mi,Binetruy:1996uv,Craig:2011yk,Brust:2011tb,Papucci:2011wy}.  In this paper, we study 
a simple, general framework in which such superparticle spectra 
with light stops are obtained naturally.

\begin{figure}[t]
\begin{center}
\begin{picture}(400,105)(0,0)
  \CArc(95,50)(45,0,360)
  \Text(96,60)[]{\large SUSY} \Line(75,55)(115,65)
  \Text(95.5,40)[]{\footnotesize (flavor universal)}
  \CArc(305,50)(45,0,360)
  \Text(306,61)[]{Electroweak} \Line(272,55)(335,63)
  \Text(305,38)[]{\large $\vev{h} \neq 0$}
  \Text(159,58)[]{\Large 1, 2} \Text(233,58)[]{\Large $q_3$, $t_R$}
  \Text(200,38)[]{\Large $\leftarrow$ $b_R$, $l_3$, $\tau_R$ $\rightarrow$}
\end{picture}
\caption{A basic scheme yielding light stop spectra.  A theory has 
 one ``dimension,'' of which electroweak and supersymmetry breakings 
 are ``located'' at the opposite ends.  This dimension may be geometric 
 or an effective one generated through dynamics.  The first two generation 
 fields are localized towards the supersymmetry breaking ``site,'' 
 obtaining flavor universal supersymmetry breaking masses and only 
 small effects from electroweak symmetry breaking (small Yukawa couplings). 
 On the other hand, top-quark multiplets are localized more towards 
 the electroweak breaking ``site,'' obtaining a large Yukawa coupling 
 but only small supersymmetry breaking effects.}
\label{fig:setup}
\end{center}
\end{figure}
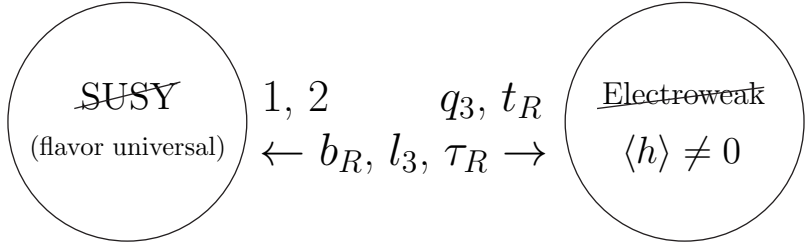

One strategy to yield such light stop spectra is to arrange the 
theory in such a way that being ``away'' from electroweak 
symmetry breaking necessarily means being ``close'' to supersymmetry 
breaking, and vice versa.  This makes the lighter generations 
(particles feeling smaller effects from electroweak symmetry breaking) 
obtain larger supersymmetry breaking masses, e.g.\ of order a few TeV, 
while keeping stops light.  Strong constraints from flavor violation 
still require the first two generation sfermions to be flavor universal, 
but this can be achieved if these generations are both strongly 
localized to the supersymmetry breaking ``site,'' and if mediation 
of supersymmetry breaking there is flavor universal.  The setup 
described here is depicted schematically in Fig.~\ref{fig:setup}.

A simple way to realize the above setup is through geometry.  Suppose 
there is an extra dimension compactified on an interval, of which 
the Higgs and supersymmetry breaking fields $h$ and $X$ are localized 
at the opposite ends.  The SM gauge, quark, and lepton multiplets 
propagate in the bulk.  Now, if two generations are localized towards 
the ``$X$ brane'' and (at least the quark doublet and up-type quark of) 
the other generation is localized towards the ``$h$ brane,'' then it 
explains the (anti-)correlation between the spectrum of SM matter and 
its superpartners---the hierarchy of the Yukawa couplings are generated 
through the wavefunction overlap of SM matter with the $h$ brane, while 
only the first two generation sfermions obtain significant supersymmetry 
breaking masses through interactions with the $X$ brane.

Another manifestation of this is through dynamics---the ``dimension'' 
separating two breakings in Fig.~\ref{fig:setup} may be generated 
effectively as a result of strong (quasi-)conformal dynamics.  Suppose 
there are elementary as well as composite sectors.  In this case, 
particles in each sector interact with significant strength, but 
interactions involving both elementary and composite particles are 
suppressed by higher dimensions of composite fields.  This can therefore 
be used to realize our setup, for example, by considering $X$ and $h$ 
to be elementary and composite fields, respectively.  The SM matter 
fields are mixture of elementary and composite states---two generations 
being mostly elementary while the other mostly composite.  In this way, 
the required pattern for the sfermion masses, as well as the hierarchical 
structure of the Yukawa couplings, are obtained.  In fact, this picture 
can be related with the geometric picture described above.  Since the 
strong, composite sector exhibits (approximate) conformality at high 
energies, the dynamics is well described by a warped extra dimension, 
using the AdS/CFT correspondence.  (For applications of this idea in other 
contexts, see e.g.~\cite{Contino:2003ve,Erlich:2005qh,Birkedal:2004zx}.)

In this paper, we present a class of models formulated in warped space, 
which can be interpreted either as a geometric or dynamical realization 
described above.  In the next section, we present the basic structure 
of the models and interpret them as composite Higgs-top models in the 
desert.  We pay particular attention to how strong constraints from 
flavor violation are avoided while generating the Yukawa hierarchy. 
In Section~\ref{sec:ewsb}, we analyze electroweak symmetry breaking and 
present sample superparticle spectra; we also give some useful formulae 
for the Higgs boson mass in the appendix.  In section~\ref{sec:flat}, 
we mention a realization of our scheme in a flat space extra dimension. 
We conclude in Section~\ref{sec:concl}.

The configuration of supersymmetry breaking and matter/Higgs fields 
in our models is the same as that in ``emergent supersymmetry'' models 
considered before~\cite{Gherghetta:2003wm,Sundrum:2009gv,Gherghetta:2011wc}, 
where the masses of elementary superpartners $\tilde{m}$ are taken (much) 
above the scale of strong dynamics $k'$.  In this picture, the quadratic 
divergence of the Higgs mass-squared parameter is regulated by a 
combination of composite Higgsinos/stops as well as higher resonances 
of the strong sector (Kaluza-Klein towers).  Instead, our picture 
here is that the theory below the compositeness scale is the full 
supersymmetric standard model, $\tilde{m} < k'$, so that the quadratic 
divergence of the Higgs mass-squared is regulated by superpartner 
loops as in usual supersymmetric models---the strong sector simply 
plays a role of generating a light stop spectrum at some energy 
$k'$.  This alleviates the problem of a potentially large $D$-term 
operator~\cite{Sundrum:2009gv}, intrinsic to the framework of 
Ref.~\cite{Gherghetta:2003wm,Sundrum:2009gv,Gherghetta:2011wc}.

Three interesting papers have recently considered light stops 
in supersymmetry~\cite{Csaki:2012fh,Craig:2012yd,Krippendorf:2012ir}, 
which are related to our study here.  Ref.~\cite{Csaki:2012fh} discusses 
supersymmetric models in which the Higgs, top, and electroweak gauge 
fields are (partial) composites of a strong sector that sits at the 
bottom edge of the conformal window.  This can be viewed as an explicit 
4D realization of our warped 5D setup.  (This ``analogy'' has also 
been drawn in that paper.)  Ref.~\cite{Craig:2012yd} considers the 
scheme of flavor mediation, where supersymmetry breaking is mediated 
through a gauged subgroup of SM flavor symmetries, leading to degenerate 
light-generation sfermions with light stops.  Ref.~\cite{Krippendorf:2012ir} 
discusses light stops in the context of heterotic string theory.

\section{Formulation in Warped Space}
\label{sec:model}

\subsection{The basic structure}
\label{subsec:basic}

In this section we present a class of models realizing the basic setup 
of Fig.~\ref{fig:setup}.  We formulate it in a 5D warped spacetime with 
the extra dimension $y$ compactified on an $S^1/Z_2$ orbifold: $0 \leq 
y \leq \pi R$.  The spacetime metric is given by
\begin{equation}
  d s^2 = e^{-2ky} \eta_{\mu\nu} dx^\mu dx^\nu + dy^2,
\label{eq:metric}
\end{equation}
where $k$ is the AdS curvature, which is taken to be somewhat (typically 
a factor of a few) smaller than the 5D cutoff scale $M_*$.  The 4D Planck 
scale, $M_{\rm Pl}$, is given by $M_{\rm Pl}^2 \simeq M_5^3/k$, where 
$M_5$ is the 5D Planck scale, and we take $k \sim M_* \sim M_5 \sim 
M_{\rm Pl}$.  For now, we take the size of the extra dimension $R$ 
to be a free parameter, satisfying $kR \simgt 1$.  If we choose $kR 
\sim 10$, the TeV scale is generated by the AdS warp factor: $k' \equiv 
k e^{-\pi kR} \sim {\rm TeV}$~\cite{Randall:1999ee}.

We consider that the SM gauge supermultiplets $\{ V_A, \Sigma_A \}$ 
($A=1,2,3$) as well as matter supermultiplets $\{ \Psi_i, \Psi^c_i \}$ 
($\Psi=Q,U,D,L,E$ with $i=1,2,3$ the generation index) propagate in 
the 5D bulk.  (Here, we have used the 4D $N=1$ superfield notation; 
see e.g.~\cite{Marti:2001iw}.) Assuming the boundary conditions
\begin{equation}
  \left( \begin{array}{c}
    V_A(+,+) \\ \Sigma_A(-,-)
  \end{array} \right),
\qquad
  \left( \begin{array}{c}
    \Psi_i(+,+) \\ \Psi^c_i(-,-)
  \end{array} \right),
\label{eq:b-c}
\end{equation}
the low-energy field content below the Kaluza-Klein (KK) excitation 
scale $\sim k'$ is the gauge and matter fields of the Minimal Supersymmetric 
Standard Model (MSSM).  They arise from the zero modes of $V_A$ and $\Psi_i$.

Now, suppose the supersymmetry breaking chiral superfield $X$ is 
localized on the ultraviolet (UV) brane at $y=0$, while two Higgs-doublet 
chiral superfields $H_u$ and $H_d$ are on the infrared (IR) brane at 
$y = \pi R$.  Then the bulk matter and gauge fields can interact with 
these fields through
\begin{equation}
  {\cal L} = \delta(y)\! 
    \biggl[ \int\! d^4\theta\, \sum_{\Psi} 
    \Bigl\{ \hat{\eta}^\Psi_{ij} X^\dagger X \Psi_i^\dagger \Psi_j 
      + \bigl( \hat{\zeta}^\Psi_{ij} X \Psi_i^\dagger \Psi_j + {\rm h.c.} 
    \bigr) \Bigr\}
  + \sum_A \Bigl\{ \int\! d^2\theta\, \hat{\xi}_A X 
      {\cal W}_A^\alpha {\cal W}_{A\alpha} + {\rm h.c.} 
    \Bigr\} \biggr]
\label{eq:susy-0}
\end{equation}
and
\begin{equation}
  {\cal L} = \delta(y-\pi R)\, e^{-3\pi kR}\!\! 
    \int\! d^2\theta\, \bigl( \hat{y}^u_{ij} Q_i U_j H_u 
      + \hat{y}^d_{ij} Q_i D_j H_d + \hat{y}^e_{ij} L_i E_j H_d 
    \bigr) + {\rm h.c.},
\label{eq:Yukawa-piR}
\end{equation}
respectively, where ${\cal W}_{A\alpha}$ are the field-strength superfields.%
\footnote{We adopt the definition of the delta function $\int_0^\varepsilon 
 \delta(y) = \int_{\pi R-\varepsilon}^{\pi R} \delta(y-\pi R) = 1$, 
 where $0 < \varepsilon< \pi R$.}

In addition, we can introduce a singlet field $S$ either in the bulk or 
on the $y=\pi R$ brane with interactions
\begin{eqnarray}
  {\cal L} &=& \delta(y-\pi R)\, e^{-3\pi kR} 
    \Bigl\{ \int\! d^2\theta\, \bigl( \hat{\lambda} S H_u H_d + \hat{f}(S) 
    \bigr) + {\rm h.c.} \Bigr\}
\nonumber\\
  && {} + \delta(y)\! \int\! d^4\theta\, 
    \Bigl\{ \hat{\eta}^S X^\dagger X S^\dagger S 
      + \bigl( \hat{\zeta}^S X S^\dagger S + {\rm h.c.} \bigr) \Bigr\},
\label{eq:singlet-int}
\end{eqnarray}
where $\hat{f}(S)$ is a holomorphic function of $S$, and the terms in 
the second line exist only if $S$ is the bulk field, $\{ S, S^c \}$. 
The introduction of $S$ allows us to accommodate a relatively heavy 
Higgs boson, including $m_h = 125~{\rm GeV}$.

The Lagrangian for the free part of a bulk supermultiplet $\{ \Phi, 
\Phi^c \}$ is given by
\begin{eqnarray}
  {\cal L} &=& e^{-2ky}\!\! \int\!d^4\theta\, 
      (\Phi^\dagger \Phi + \Phi^c \Phi^{c\dagger}) 
    + e^{-3ky} \biggl\{ \int\! d^2\theta\, \Phi^c \Bigl( \partial_y 
      + M_\Phi - \frac{3}{2}k \Bigr) \Phi + {\rm h.c.} \biggr\}
\nonumber\\
  && {} + \delta(y) \int\!d^4\theta\, z_\Phi \Phi^\dagger \Phi,
\label{eq:free-Lag}
\end{eqnarray}
where we have included a UV-brane localized kinetic term $z_\Phi$ ($>0$), 
which plays an important role in our discussion.  (A possible IR-brane 
localized kinetic term is irrelevant for the discussion.)  There are 
two parameters in this Lagrangian: $M_\Phi$ and $z_\Phi$.  The parameter 
$M_\Phi$ controls the wavefunction profile of the zero mode in the bulk. 
For $M_\Phi > k/2$ ($< k/2$) the wavefunction of a zero mode arising from 
$\Phi$ is localized to the UV (IR) brane; for $M_\Phi = k/2$ it is flat 
(see e.g.~\cite{Gherghetta:2000qt}).  The parameter $z_\Phi$ is important 
for a field with $M_\Phi \simgt k/2$; it controls how much of the zero 
mode is regarded as the brane and bulk degrees of freedom.  For $z_\Phi 
M_\Phi \gg 1$, the zero mode is mostly brane field-like, while for 
$z_\Phi M_\Phi \ll 1$ it is bulk field-like.

Our setup is realized by taking $M_\Phi \simgt k/2$ and $z_\Phi M_\Phi 
\gg 1$ for the first two generations of matter while $M_\Phi \ll k/2$ 
for the third generation quark-doublet and right-handed top multiplets 
$\{ Q_3, Q_3^c \}$ and $\{ U_3, U_3^c \}$.  This implies that the former 
are mostly brane field-like, while the latter are bulk fields with the 
wavefunctions localized to the IR brane.  (In the 4D interpretation 
discussed in Section~\ref{subsec:4D-interp}, these correspond to 
mostly elementary and composite fields, respectively.)  The zero-mode 
wavefunctions for the other third generation multiplets $\{ D_3, D_3^c \}$, 
$\{ L_3, L_3^c \}$, and $\{ E_3, E_3^c \}$ are more flexible, although 
they are still subject to constraints from flavor physics, both to 
reproduce realistic Yukawa matrices and to avoid excessive supersymmetric 
contributions to flavor violation.

More specifically, the wavefunction of the zero mode of the $\{ \Phi, 
\Phi^c \}$ multiplet in Eq.~(\ref{eq:free-Lag}) is given by
\begin{equation}
  f_\Phi(y) = \frac{1}{\sqrt{z_\Phi + \frac{1}{2(M_\Phi - \frac{k}{2})} 
    (1-e^{-2\pi R(M_\Phi-\frac{k}{2})})}}\, e^{-(M_\Phi-\frac{k}{2})y}
\label{eq:wavef}
\end{equation}
in the ``conformal-field'' basis, in which 5D scalar and fermion fields 
$\phi$ and $\psi$ are rescaled from the original component fields in $\Phi$ 
as $\phi = e^{-ky} \Phi|_0$ and $\psi_\alpha = e^{-ky} \Phi|_{\theta}$. 
The low-energy 4D theory below $\sim k'$ is obtained by integrating over 
$y$ with this wavefunction.  For the superpotential terms, it leads to
\begin{equation}
  {\cal L}_{\rm 4D} = \int\! d^2\theta\, \bigl( y^u_{ij} Q_i U_j H_u 
      + y^d_{ij} Q_i D_j H_d + y^e_{ij} L_i E_j H_d 
      + \lambda S H_u H_d + f(S) \bigr) + {\rm h.c.},
\label{eq:superpot-4D}
\end{equation}
where the 4D coupling constants (quantities without hat) are related 
with the 5D ones (with hat) by
\begin{equation}
  y^u_{ij} = \hat{y}^u_{ij}\, x_{Q_i} x_{U_j},
\qquad
  y^d_{ij} = \hat{y}^d_{ij}\, x_{Q_i} x_{D_j},
\qquad
  y^e_{ij} = \hat{y}^e_{ij}\, x_{L_i} x_{E_j},
\qquad
  \lambda = \hat{\lambda}\, x_S,
\label{eq:4D-Yukawa}
\end{equation}
and $f(S) = \hat{f}(x_S S)$.  Here, the factors $x_\Phi$ ($\Phi = Q_i, 
U_i, D_i, L_i, E_i, S$) are given by
\begin{equation}
  x_\Phi \equiv f_\Phi(\pi R)
  \simeq \left\{ 
  \begin{array}{ll}
    \frac{1}{\sqrt{z_\Phi + \frac{1}{2M_\Phi}}}\, e^{-\pi R M_\Phi} & 
      {\rm for}\,\,\, M_\Phi \gg \frac{k}{2}
\\
    \frac{1}{\sqrt{z_\Phi + \pi R}} & 
      {\rm for}\,\,\, M_\Phi \sim \frac{k}{2}
\\
    \sqrt{2(\frac{k}{2} - M_\Phi)} & 
      {\rm for}\,\,\, M_\Phi \ll \frac{k}{2}
  \end{array}, \right.
\label{eq:epsilon}
\end{equation}
where we have used $z_\Phi k e^{-\pi kR} \ll 1$ for $M_\Phi \ll k/2$, 
which is satisfied in the relevant parameter region considered later.%
\footnote{The expression $M_\Phi \ll k/2$ here and after means that 
 $|M_\Phi| \ll k/2$ or $M_\Phi < 0$.}
The case with brane $S$ is obtained by replacing $x_S$ with $1$.

For the supersymmetry-breaking terms, the 4D theory below $\sim k'$ yields
\begin{eqnarray}
  {\cal L}_{\rm 4D} &=& \int\! d^4\theta\, 
    \Bigl\{ \sum_{\Psi} \eta^\Psi_{ij} X^\dagger X \Psi_i^\dagger \Psi_j 
      + \eta^S X^\dagger X S^\dagger S 
    + \bigl( \sum_{\Psi} \zeta^\Psi_{ij} X \Psi_i^\dagger \Psi_j 
      + \zeta^S X S^\dagger S + {\rm h.c.} \bigr) \Bigr\}
\nonumber\\
  && {} + \sum_A \Bigl\{ \int\! d^2\theta\, \xi_A X 
      {\cal W}_A^\alpha {\cal W}_{A\alpha} + {\rm h.c.} 
    \Bigr\},
\label{eq:susy-br-4D}
\end{eqnarray}
where
\begin{equation}
  \eta^\Psi_{ij} = \hat{\eta}^\Psi_{ij}\, r_{\Psi_i} r_{\Psi_j},
\qquad
  \eta^S = \hat{\eta}^S\, r_S^2,
\qquad
  \zeta^\Psi_{ij} = \hat{\zeta}^\Psi_{ij}\, r_{\Psi_i} r_{\Psi_j},
\qquad
  \zeta^S = \hat{\zeta}^S\, r_S^2,
\qquad
  \xi_A = \hat{\xi}_A,
\label{eq:4D-susybr}
\end{equation}
and the factors $r_\Phi$ are given by
\begin{equation}
  r_\Phi \equiv f_\Phi(0) 
  \simeq \left\{ 
  \begin{array}{ll}
    \frac{1}{\sqrt{z_\Phi + \frac{1}{2M_\Phi}}} & 
      {\rm for}\,\,\, M_\Phi \gg \frac{k}{2}
\\
    \frac{1}{\sqrt{z_\Phi + \pi R}} & 
      {\rm for}\,\,\, M_\Phi \sim \frac{k}{2}
\\
    \sqrt{2(\frac{k}{2} - M_\Phi)}\, e^{-\pi R(\frac{k}{2}-M_\Phi)} & 
      {\rm for}\,\,\, M_\Phi \ll \frac{k}{2}
  \end{array}. \right.
\label{eq:r}
\end{equation}
The case with brane $S$ is obtained by $r_S \rightarrow 0$.

As will be discussed in Section~\ref{subsec:4D-interp}, the models 
presented here can be interpreted, through the AdS/CFT correspondence, 
as those of composite Higgs-top in the supersymmetric desert.  As 
such, small neutrino masses can be generated by the conventional 
seesaw mechanism.  Specifically, we can introduce right-handed neutrino 
supermultiplets $\{ N_i, N_i^c \}$ in the bulk, with Majorana masses and 
neutrino Yukawa couplings located on the UV and IR branes, respectively:
\begin{equation}
  {\cal L} = \delta(y)\! 
      \int\! d^2\theta\, \frac{\hat{M}_{ij}}{2} N_i N_j 
    + \delta(y-\pi R)\, e^{-3\pi kR}\!\! 
      \int\! d^2\theta\, \hat{y}^\nu_{ij} L_i N_j H_u 
    + {\rm h.c.}
\label{eq:neutrino}
\end{equation}
For $M_{N_i} \sim k/2$, this naturally generates small neutrino 
masses of the observed size (assuming the absence of tree-level 
neutrino-mass operators such as $\int\! d^2\theta\, (LH_u)^2$ on the 
IR brane)~\cite{Nomura:2003du}.  Alternatively, small Dirac neutrino 
masses can be obtained if we prohibit the Majorana masses for $N_i$ 
and localize them to the UV brane~\cite{Grossman:1999ra}.

\subsection{Physics of flavor---fermions and sfermions}
\label{subsec:flavor}

We now discuss the flavor structure of quarks/leptons and squarks/sleptons 
in more detail.  Suppose that all the couplings on the UV brane are 
roughly of $O(1)$ in units of some messenger scale $M_{\rm mess}$.  In 
this case, Eqs.~(\ref{eq:4D-susybr},~\ref{eq:r}) imply that the zero 
modes localized to the IR brane obtain only exponentially suppressed 
supersymmetry-breaking masses (at scale $k'$):
\begin{equation}
  m_{\tilde{Q}_3,\tilde{U}_3} / m_{\tilde{\Psi}_{1,2}} \ll 1.
\label{eq:light-stop}
\end{equation}
A main motivation to consider light stops is naturalness, 
Eq.~(\ref{eq:stop}).  To keep this, we take $m_{\tilde{Q}_3,\tilde{U}_3} 
\simlt (400~\mbox{--}~500)~{\rm GeV}$ (after evolving down to the weak 
scale).  In order to satisfy constraints from flavor violation, the 
right-handed bottom and first two generation squark masses should 
be in the multi-TeV region~\cite{Brust:2011tb,Giudice:2008uk}. 
We therefore choose $M_{D_3} \simgt k/2$, and
\begin{eqnarray}
  && m_{\tilde{\Psi}_{1,2}} \sim m_{\tilde{b}_R} \sim \mbox{a few TeV},
\label{eq:heavy-sf}\\
  && m_{\tilde{t}_L,\tilde{t}_R} \sim m_{\tilde{b}_L} 
    \simlt (400~\mbox{--}~500)~{\rm GeV}.
\label{eq:light-sf}
\end{eqnarray}
The masses of $\tilde{L}_3$ and $\tilde{E}_3$ are less constrained, 
although we consider $M_{L_3,E_3} \simgt k/2$ in most of the 
paper, leading to $m_{\tilde{\tau}_L,\tilde{\tau}_R,\tilde{\nu}_\tau} 
\sim \mbox{a few TeV}$.  With the mass splitting of 
Eqs.~(\ref{eq:heavy-sf},~\ref{eq:light-sf}), the hypercharge 
$D$-term contribution does not have a large effect on the Higgs 
mass-squared parameter to destabilize naturalness.

The masses of the gauginos are determined by parameters such as 
$\hat{\xi}_A$, $\hat{\eta}^\Psi_{ij}$ and $z_\Phi$, which depend on 
a detailed mechanism generating operators in Eq.~(\ref{eq:susy-0}). 
Motivated by naturalness, in this paper we take
\begin{equation}
  m_{\tilde{B},\tilde{W}} \simlt 1~{\rm TeV},
\quad
  m_{\tilde{g}} \sim 1~{\rm TeV}.
\label{eq:gaugeinos}
\end{equation}
The gluino mass, $m_{\tilde{g}} \simeq M_3$, is chosen so that the stops 
do not obtain large radiative corrections exceeding Eq.~(\ref{eq:light-sf}), 
and that the theory is not excluded by the LHC data: $m_{\tilde{g}} 
\simgt 700~{\rm GeV}$~\cite{Papucci:2011wy}.  The above equations 
(\ref{eq:heavy-sf}~--~\ref{eq:gaugeinos}) specify the superpartner 
spectra we consider.%
\footnote{In deriving these expressions, we have ignored possible 
 contributions to the supersymmetry breaking masses from the sector 
 that stabilizes the radius of the extra dimension.  This assumption 
 is justified for certain ways of stabilizing the radius; see, e.g., 
 Ref.~\cite{Gherghetta:2011wc}.}

What about the flavor structure for quarks/leptons and those among the 
first two generation sfermions?  In this paper, we consider a theory in 
which all the nontrivial flavor structures are generated from physics 
of the bulk (and on the IR brane).  In the 4D ``dual'' picture discussed 
in Section~\ref{subsec:4D-interp}, this corresponds to the setup in which 
the nontrivial flavor structure is generated through interactions of the 
elementary sector with the strongly-interacting composite sector.  This 
implies that all the flavor violating effects are shut off in the high 
energy limit, giving the conditions
\begin{equation}
  \hat{\eta}^\Psi_{ij} \propto \hat{\zeta}^\Psi_{ij} \propto (z_\Psi)_{ij}
\label{eq:UV}
\end{equation}
in flavor space, $i,j = 1,2,3$.  In particular, in the field basis 
that $(z_\Psi)_{ij} \propto \delta_{ij}$, which we can always take, 
$\hat{\eta}^\Psi_{ij} \propto \hat{\zeta}^\Psi_{ij} \propto \delta_{ij}$. 
This can be achieved if the operators in Eq.~(\ref{eq:susy-0}) are 
generated by flavor universal dynamics, e.g.\ gauge mediation on 
the UV brane.

With the multi-TeV masses, the spectrum of the first two generation 
sfermions must be somewhat degenerate, to avoid stringent constraints 
from flavor.  From Eq.~(\ref{eq:r}), we find that the first two generation 
sfermion masses depend on $\hat{\eta}^\Psi_{ij}$, $(z_\Psi)_{ij}$, 
and $(M_\Psi)_{ij}$.  (Note that we take the bulk masses larger than 
$k/2$ for the first two generations of matter.)  In the field basis 
that $\hat{\eta}^\Psi_{ij}$ and $(z_\Psi)_{ij}$ are proportional to the 
unit matrix, $\hat{\eta}^\Psi_{ij} \equiv \hat{\eta}^\Psi \delta_{ij}$ 
and $(z_\Psi)_{ij} \equiv z_\Psi \delta_{ij}$, the only source of flavor 
violation comes from $(M_\Psi)_{ij}$, which we can diagonalize by field 
rotation in flavor space: $(M_\Psi)_{ij} = M_{\Psi_i} \delta_{ij}$. 
The effects of flavor violation are then of order
\begin{equation}
  \frac{\varDelta \tilde{m}_{ij}}{\tilde{m}_i+\tilde{m}_j} 
  = \frac{r_{\Psi_i} - r_{\Psi_j}}{r_{\Psi_i} + r_{\Psi_j}},
\label{eq:12-split}
\end{equation}
multiplied by appropriate flavor mixing angles arising from diagonalization 
of the 4D Yukawa matrices.  Here, $r_{\Psi_i}$ are given in Eq.~(\ref{eq:r}). 
Requiring that these effects satisfy constraints from the $K$-$\bar{K}$ 
physics~\cite{Gabbiani:1996hi}, we find, for example,
\begin{equation}
  z_\Psi k \simgt \{ 15, 12, 4 \}
\quad
  \mbox{for  } \tilde{m} = \{ 1, 4, 10 \}~{\rm TeV},
\label{eq:z_Psi}
\end{equation}
for $M_{\Psi_2}/k \simeq 0.6$ and $M_{\Psi_1}/k \simeq 0.7$, which 
produces hierarchy of $O(0.1)$ by the difference of wavefunction profiles 
between $\Psi_1$ and $\Psi_2$.  Here, $\tilde{m}$ represents the masses 
of the first two generation sfermions, and we have assumed the maximal 
phase in the relevant matrix element.  While the precise constraint 
on $z_\Psi$ depends on detailed modeling of flavor, we generically 
need nonvanishing $z_\Psi \simgt O(10/k)$ in the case of the maximal 
phase in $K$-$\bar{K}$ mixing.%
\footnote{In the 4D picture of Section~\ref{subsec:4D-interp}, this 
 corresponds to the case where the first two generations of matter 
 are mostly elementary, with the contributions of the strong sector 
 to their kinetic terms suppressed compared to those at tree level.}

The structure of the 4D Yukawa couplings can be read off from 
Eqs.~(\ref{eq:4D-Yukawa},~\ref{eq:epsilon}).  For a field with $M_\Phi > 
k/2$, we have a suppression arising from the wavefunction profile of the 
zero mode, $\epsilon_\Phi \equiv e^{-\pi R (M_\Phi-k/2)}$, contributing to 
the hierarchy of the Yukawa couplings~\cite{Gherghetta:2000qt,Hamidi:1986vh}. 
In addition, fields with $M_{\Psi_{1,2}} \simgt k/2$ may have an additional 
suppression $\epsilon \equiv 1/\sqrt{z_\Phi M_*}$ if $z_\Phi k \gg 1$. 
For example, if we take $M_{D_3,L_3,E_3} \simgt k/2$, then we find
\begin{equation}
  y_u \simprop
  \left( \begin{array}{ccc}
    \epsilon^2 \epsilon_{Q_1} \epsilon_{U_1} & 
      \epsilon^2 \epsilon_{Q_1} \epsilon_{U_2} & \epsilon\, \epsilon_{Q_1} \\
    \epsilon^2 \epsilon_{Q_2} \epsilon_{U_1} & 
      \epsilon^2 \epsilon_{Q_2} \epsilon_{U_2} & \epsilon\, \epsilon_{Q_2} \\
    \epsilon\, \epsilon_{U_1} & \epsilon\, \epsilon_{U_2} & 1 
  \end{array} \right),
\qquad
  y_d \simprop
  \left( \begin{array}{ccc}
    \epsilon^2 \epsilon_{Q_1} \epsilon_{D_1} & 
      \epsilon^2 \epsilon_{Q_1} \epsilon_{D_2} & 
      \epsilon^2 \epsilon_{Q_1} \epsilon_{D_3} \\
    \epsilon^2 \epsilon_{Q_2} \epsilon_{D_1} & 
      \epsilon^2 \epsilon_{Q_2} \epsilon_{D_2} & 
      \epsilon^2 \epsilon_{Q_2} \epsilon_{D_3} \\
    \epsilon\, \epsilon_{D_1} & \epsilon\, \epsilon_{D_2} & 
      \epsilon\, \epsilon_{D_3} 
  \end{array} \right),
\label{eq:y_ud}
\end{equation}
\begin{equation}
  y_e \simprop
  \left( \begin{array}{ccc}
    \epsilon^2 \epsilon_{L_1} \epsilon_{E_1} & 
      \epsilon^2 \epsilon_{L_1} \epsilon_{E_2} & 
      \epsilon^2 \epsilon_{L_1} \epsilon_{E_3} \\
    \epsilon^2 \epsilon_{L_2} \epsilon_{E_1} & 
      \epsilon^2 \epsilon_{L_2} \epsilon_{E_2} & 
      \epsilon^2 \epsilon_{L_2} \epsilon_{E_3} \\
    \epsilon^2 \epsilon_{L_3} \epsilon_{E_1} & 
      \epsilon^2 \epsilon_{L_3} \epsilon_{E_2} & 
      \epsilon^2 \epsilon_{L_3} \epsilon_{E_3} 
  \end{array} \right),
\label{eq:y_e}
\end{equation}
where $O(1)$ factors are omitted in each element, and $\epsilon_\Phi 
\ll 1$ only if $\pi R(M_\Phi - k/2) \gg 1$ and $\epsilon \ll 1$ only if 
$z_\Phi k \gg 1$.  Therefore, with suitable choices for $M_{\Psi_i}$, 
the observed pattern of the Yukawa couplings can be reproduced through 
physics of the bulk (i.e.\ the dynamics of the strong sector in the 4D 
picture) while keeping approximate flavor universality for the first 
two generation sfermion masses.

\subsection{4D interpretation}
\label{subsec:4D-interp}

Models discussed here can be interpreted as purely 4D models 
formulated in the conventional grand desert, using the AdS/CFT 
correspondence.  (For discussions on this correspondence, see 
e.g.~\cite{Contino:2003ve,ArkaniHamed:2000ds}.)  In the 4D picture, 
the first two generations of matter are (mostly) elementary, while the 
third generation quark-doublet and right-handed top multiplets arise 
as composite fields of some strongly interacting sector, which exhibits 
nontrivial dynamics at an exponentially small scale $\approx k' = 
k e^{-\pi kR}$.  (We mostly consider that the right-handed bottom 
and third-generation lepton multiplets are elementary, although there 
is some flexibility on this choice.)  This strong dynamics also produces 
$S$, $H_u$, and $H_d$ fields, together with superpotential interactions 
$W_H = \lambda S H_u H_d + f(S)$ at $k'$.  (We focus on the case of 
IR-brane localized $S$ in this section.)  Since the Higgs-top sector 
is strongly coupled at $k'$, the Landau pole constraint for the 
couplings in $W_H$ (and the top Yukawa coupling) needs to be satisfied 
only below $k'$~\cite{Birkedal:2004zx}, realizing the $\lambda$SUSY 
framework in Ref.~\cite{Barbieri:2006bg}.

Supersymmetry breaking is mediated at the scale $M_{\rm mess}$, giving 
TeV to multi-TeV masses to the elementary sfermions as well as the gauginos. 
The effect of supersymmetry breaking in the composite sector is diluted 
by the near-conformal strong dynamics~\cite{Gherghetta:2003wm}, as 
long as operators associated with this effect have large anomalous 
dimensions~\cite{Sundrum:2009gv}.  This therefore yields only negligible 
soft masses for the composite fields at $k'$.%
\footnote{In our models, supersymmetry breaking masses in the 
 elementary sector, $\tilde{m} \sim \mbox{a few TeV}$, is smaller 
 than the compositeness scale, $k' \simgt 10~{\rm TeV}$ (see 
 below).  Therefore, the problem of a potentially large $D$-term 
 operator~\cite{Sundrum:2009gv}, intrinsic to the framework of 
 Refs.~\cite{Gherghetta:2003wm,Sundrum:2009gv,Gherghetta:2011wc}, 
 does not arise, unless this operator is generated directly by the 
 physics at $M_{\rm mess}$.  The dilution of supersymmetry breaking 
 effects in the composite sector has been studied explicitly in 
 Ref.~\cite{Csaki:2012fh} in a setup similar to ours, using 
 Seiberg duality.}
A composite field, however, may obtain sizable supersymmetry breaking 
masses (only) if it mixes with an elementary state, which in the 5D 
picture corresponds to delocalizing the state from the IR brane.

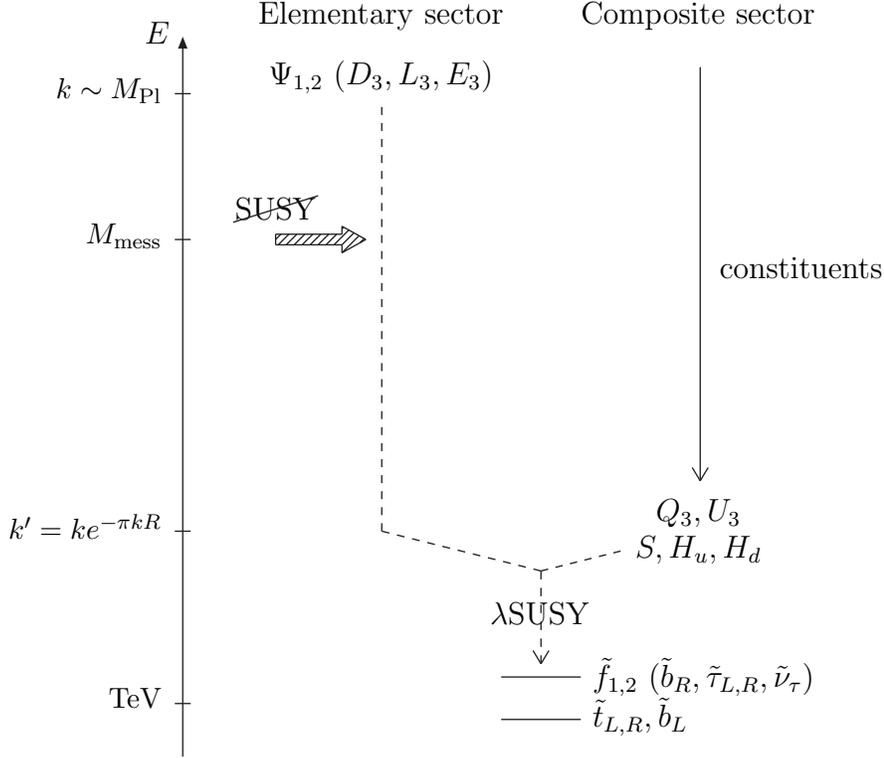
\begin{figure}[t]
\begin{center}
\begin{picture}(330,300)(-50,-45)
  \LongArrow(0,-40)(0,230) \Text(-5,230)[br]{$E$}
  \Line(-3,210)(3,210) \Text(-8,212)[r]{\small $k \sim M_{\rm Pl}$}
  \Line(-3,155)(3,155) \Text(-8,157)[r]{\small $M_{\rm mess}$}
  \Line(-3,45)(3,45) \Text(-8,47)[r]{\small $k' = k e^{-\pi kR}$}
  \Line(-3,-20)(3,-20) \Text(-8,-18)[r]{\small TeV}
  \Text(75,235)[b]{Elementary sector}
  \Text(75,217)[]{$\Psi_{1,2}$ ($D_3, L_3, E_3$)}
  \DashLine(75,205)(75,45){3}
  \Text(35,163)[b]{SUSY} \Line(19,161.5)(51,171.5)
  \Line(35,153)(35,157) \Line(35,157)(60,157) \Line(35,153)(60,153)
  \Line(60,157)(60,160) \Line(60,160)(69,155)
  \Line(60,153)(60,150) \Line(60,150)(69,155)
  \Line(35,155)(37,157) \Line(36,153)(40,157) \Line(39,153)(43,157)
  \Line(42,153)(46,157) \Line(45,153)(49,157) \Line(48,153)(52,157)
  \Line(51,153)(55,157) \Line(54,153)(58,157) \Line(60,159)(60.6,159.6)
  \Line(57,153)(62.7,158.7) \Line(60,153)(64.6,157.6)
  \Line(60,150)(66.5,156.5) \Line(66.8,153.8)(68.4,155.4)
  \Text(195,235)[b]{Composite sector}
  \Line(195,220)(195,64) \Text(203,145)[l]{constituents}
  \Line(195,64)(192,69) \Line(195,64)(198,69)
  \Text(195,52)[]{$Q_3, U_3$}
  \Text(195,38)[]{$S, H_u, H_d$}
  \DashLine(75,45)(135,30){3} \DashLine(165,37.5)(135,30){3}
  \DashLine(135,30)(135,-5){3} \Text(135,14)[]{$\lambda$SUSY}
  \Line(135,-5)(132,0) \Line(135,-5)(138,0)
  \Line(120,-10)(150,-10) \Text(155,-10)[l]{$\tilde{f}_{1,2}$ 
    ($\tilde{b}_R, \tilde{\tau}_{L,R}, \tilde{\nu}_\tau$)}
  \Line(120,-26)(150,-26) \Text(155,-25)[l]{$\tilde{t}_{L,R}, \tilde{b}_L$}
\end{picture}
\caption{The 4D interpretation of the models.  The first two generations 
 of matter are elementary, while the Higgs and top multiplets arise as 
 composite states at the scale of strong dynamics, $k'$.  The theory 
 between $k'$ and TeV is $\lambda$SUSY in Ref.~\cite{Barbieri:2006bg} 
 (with a modest Higgs-sector coupling $\lambda$), with light stops and 
 left-handed sbottom.}
\label{fig:4D}
\end{center}
\end{figure}

The top Yukawa coupling is naturally large as the relevant fields are 
all composite.  On the other hand, the Yukawa couplings for the first 
two generations of matter are generated through mixing of these states 
with fields in the composite sector, so are suppressed.  The amount 
of suppression depends on the dimension of the mixing operator, and 
thus varies field by field, yielding a hierarchical pattern for the 
Yukawa matrices.  Note that this way of dynamically generating the Yukawa 
hierarchy does not contradict the stringent constraints on supersymmetric 
flavor violation as long as supersymmetry breaking mediation at 
$M_{\rm mess}$ is flavor universal (e.g.\ as in the case of gauge 
mediation) {\it and} the contribution to the kinetic terms of the 
elementary fields from the strong sector is small (which corresponds 
to the condition in Eq.~(\ref{eq:z_Psi}) in 5D).  The overall picture 
for the 4D interpretation described here is depicted schematically 
in Fig.~\ref{fig:4D}.

The value of the compositeness scale $k'$ is constrained by phenomenological 
considerations.  As in Eq.~(\ref{eq:light-sf}), we take the stops 
light to keep electroweak symmetry breaking natural.  On the other 
hand, the LHC bound on the gluino mass for these values of stop masses 
is $m_{\tilde{g}} \simgt 700~{\rm GeV}$, so that we need a little 
``hierarchy'' between $m_{\tilde{t}}$ and $m_{\tilde{g}}$.  Since 
$m_{\tilde{t}}$ receives a positive contribution from $m_{\tilde{g}}$ 
through renormalization group evolution, this bounds the scale $k'$ 
from above.  The precise bound is (exponentially) sensitive to the low 
energy parameters, but we typically find that $k'$ must be below an 
intermediate scale; in particular, it cannot be at the unification 
scale.  The value of $k'$ is also limited from above by Landau pole 
considerations for the couplings in $W_H$.

The lower bound on $k'$ can be obtained for a fixed $m_{\tilde{g}}$ 
by requiring that $\tilde{t}_L$, $\tilde{t}_R$, and $\tilde{b}_L$ are 
sufficiently heavy to avoid the LHC bounds.  Assuming that these states 
decay either into the lightest neutralino or the gravitino within the 
detector, which we would need anyway to avoid a strong constraint on 
quasi-stable stops, the masses of $\tilde{t}_L$ and $\tilde{b}_L$ 
must be larger than about $250~{\rm GeV}$~\cite{Papucci:2011wy}. 
Moreover, if the neutralino to which these states decay is lighter 
than $\approx 100~{\rm GeV}$ (or if they decay into the gravitino), 
then the mass of $\tilde{b}_L$ must be larger than about $400~{\rm 
GeV}$~\cite{Aad:2011cw}.  Since the running masses for these states, 
$m_{\tilde{Q}_3}$ and $m_{\tilde{U}_3}$, are vanishing at $k'$ (up 
to small threshold corrections), this limits $k'$ from below for 
a fixed $m_{\tilde{g}}$.  In this paper, we take
\begin{equation}
  k' \simgt 10~{\rm TeV},
\label{eq:k'}
\end{equation}
so that the theory below the compositeness scale is the 
supersymmetric standard model with the superpartner spectrum given 
by Eqs.~(\ref{eq:heavy-sf}~--~\ref{eq:gaugeinos}).  With these values 
of $k'$, other lower bounds on $k'$ coming from precision electroweak 
measurements and flavor/$CP$ violation induced by KK excitations are 
satisfied~\cite{Davoudiasl:2009cd}.  (Note that the masses of the 
lowest KK excitations are given by $\approx \pi k'$.)

Our models have the supersymmetric grand desert between $k'$ and $k \sim 
M_{\rm Pl}$.  Thus, if the strong sector respects a (global) unified 
symmetry, then we can discuss gauge coupling unification, along the lines 
of Ref.~\cite{Goldberger:2002pc}.  The prediction depends on the location 
of matter fields, especially $D_3, L_3$ and $E_3$; in the minimal case 
where these fields have $M_\Phi \simgt k/2$, the three SM gauge couplings 
approach at $\sim 10^{17}~{\rm GeV}$, but with the precision of unification 
worse than that in the SM ($\delta g^2/\bar{g}^2 \approx 15\%$ at the 
unification scale).  We do not pursue the issue of unification further 
in this paper.

\section{Electroweak Symmetry Breaking}
\label{sec:ewsb}

\subsection{Overview}
\label{subsec:overview}

As outlined in Section~\ref{subsec:4D-interp}, our theory above the 
compositeness scale $k'$ is $SU(3)_C \otimes SU(2)_L \otimes U(1)_Y$ 
gauge theory that has the elementary fields $\Psi_{1,2}$ (and $D_3, 
L_3, E_3$) and the strongly interacting near-conformal sector.  The 
beta functions for the gauge group are given by
\begin{equation}
  b_A = b^{\rm MSSM}_A - b^{Q_3+U_3+H_{u,d}}_A + b_{\rm CFT},
\label{eq:beta-func}
\end{equation}
where $b_{\rm CFT}$ is the contribution from the strong sector, which 
corresponds to $1/g_{\rm 5D}^2 k$ in the 5D picture, and is universal 
if this sector respects a (global) unified symmetry.  Supersymmetry 
breaking masses for the elementary fields, including the gaugino masses 
$M_A$, are generated at $M_{\rm mess}$, and they are evolved down to 
$k'$ by the renormalization group equations with Eq.~(\ref{eq:beta-func}). 
The composite fields appear at $k'$, which have vanishing supersymmetry 
breaking masses at that scale (up to small threshold corrections of 
$O(M_A^2/16\pi^2)$ in squared masses).

Physics of electroweak symmetry breaking is governed by the dynamics 
of the composite sector and the gaugino masses.  At scale $k'$, the 
strong sector produces the superpotential
\begin{equation}
  W_H = \lambda S H_u H_d + f(S) + \cdots,
\label{eq:WH-general}
\end{equation}
for the Higgs sector, where the dots represent higher dimension terms 
which are generically suppressed by the warped-down cutoff scale 
$M'_* = M_* e^{-\pi kR}$.  In case $M'_*$ is close to the TeV 
scale, these higher dimension terms could affect phenomenology; 
for example, the term $(H_u H_d)^2$ can contribute to the Higgs 
boson mass~\cite{Brignole:2003cm}.  Similarly, higher dimension terms 
in the K\"{a}hler potential may affect phenomenology; for example, 
the terms $S^\dagger H_u H_d$ and $S^\dagger H_u^\dagger Q D$ can 
lead to a $\mu$ term and down-type quark masses if $S$ has an 
$F$-term expectation value.

In general, for relatively large values of $k'$ envisioned in 
Eq.~(\ref{eq:k'}), the effects of these higher dimension operators 
are insignificant, except possibly for light quark/lepton masses.  We 
therefore consider only renormalizable terms in the Higgs potential. 
In particular, in the rest of the paper we focus on the case where 
$W_H$ contains only dimensionless terms in 4D, and discuss how 
electroweak symmetry breaking can work in our models.  In doing so, 
we assume
\begin{equation}
  m_{\tilde{Q}_3, \tilde{U}_3, H_u, H_d} \approx 0,
\label{eq:m-bc}
\end{equation}
at $k'$, i.e.\ we ignore possible threshold corrections at that 
scale, which are highly model-dependent.  (We later consider dynamics 
at the IR scale in which non-vanishing $m_S^2$ is generated at $k'$ 
to reproduce correct electroweak symmetry breaking.)  This will 
illustrate basic features of electroweak symmetry breaking in our 
framework, in the minimal setup.

\subsection{Higgs sector: {\boldmath $\kappa$}SUSY}
\label{subsec:Higgs}

We consider a variant on the $\lambda$SUSY model~\cite{Barbieri:2006bg}, 
which has the superpotential of the Next-to-Minimal Supersymmetric 
Standard Model (NMSSM) form:
\begin{equation}
  W_H = \lambda S H_u H_d + \frac{\kappa}{3} S^3.
\label{eq:WH-NMSSM}
\end{equation}
To distinguish from other $\lambda$SUSY studies in which the $\kappa$ 
term does not play a dominant role, we call this model $\kappa$SUSY. 
We assume that $S$, $H_u$, and $H_d$ are all localized to the IR brane, 
so we require $\lambda$ and $\kappa$ to be perturbative only up to 
the scale $k'$, which we take to be $10~\mbox{--}~1000~{\rm TeV}$. 
For $k'=10~{\rm TeV}$, for example, we obtain $\lambda(M_Z) \simlt 
1.8$ for $\kappa(M_Z) = 0.7$; for $k' = 1000~{\rm TeV}$, $\lambda(M_Z) 
\simlt 1.1$ for $\kappa(M_Z) = 0.7$.

Because of Eq.~(\ref{eq:m-bc}), the only relevant dimensionful parameters 
for electroweak symmetry breaking are the gaugino masses, except possibly 
for the supersymmetry breaking mass for the $S$ field (which we will 
introduce in the next subsection).  They set the scale for the soft 
supersymmetry breaking masses in the scalar potential
\begin{eqnarray}
V &=& |\lambda H_u H_d + \kappa S^2|^2 + |\lambda S H_u|^2 
    + |\lambda S H_d|^2
\nonumber\\
  && {} + m_S^2 |S|^2 + m_{H_u}^2 |H_u|^2 + m_{H_d}^2 |H_d|^2 
    + (\lambda A_\lambda S H_u H_d - \frac{\kappa}{3} A_\kappa S^3 
    + {\rm h.c.})
\nonumber\\
  && {} + \frac{g^2}{8}(H_u^\dagger \sigma^a H_u + H_d^\dagger \sigma^a H_d)^2 
    + \frac{g'^2}{8} (|H_u|^2 - |H_d|^2 )^2,
\label{eq:scalar-pot}
\end{eqnarray}
through renormalization group evolution below $k'$.  Successful electroweak 
symmetry breaking requires all the $S$, $H_u$, and $H_d$ fields to obtain 
vacuum expectation values, $v_s \equiv \vev{S}$, $v_u \equiv \vev{H_u}$, 
and $v_d \equiv \vev{H_d}$.

Once the singlet has a vacuum expectation value, $v_s$, we obtain 
$\mu = \lambda v_s$ and $B_\mu = \lambda A_{\lambda} v_s + \kappa \lambda 
v_s^2 = \mu(A_{\lambda} + \kappa \mu/\lambda)$, where $B_\mu$ is the 
holomorphic supersymmetry breaking Higgs mass-squared.  We thus obtain 
the following Higgs mass-squared matrix (in the $h_u$-$h_d$-$s$ basis):
\begin{equation}
\begin{array}{rl}
  & {\cal M}_{\rm scalar}^2 
    \equiv \frac{1}{2} \frac{\partial^2 V}{\partial v_i \partial v_j} 
    = \frac{1}{2} \left(\frac{\partial^2 V}{\partial v_i \partial v_j} 
      - \delta_{ij} \frac{1}{v_i} \frac{\partial V}{\partial v_i} \right) 
    = \frac{1}{2}\times\\
&\\
  & \!\!\!\!\!\!\!\! \left( \begin{matrix} 
  \bar{g}^2 v_u^2 + \frac{2 B_\mu}{\tan\beta} & 
  (4\lambda^2 - \bar{g}^2) v_u v_d - 2 B_\mu & 
  4\mu v_u \Bigl( \lambda - \frac{\kappa}{\tan\beta} 
    - \frac{\lambda A_\lambda}{2\mu \tan\beta} \Bigr)
\\
  (4\lambda^2 - \bar{g}^2) v_u v_d - 2 B_\mu & 
  \bar{g}^2 v_d^2 + 2 B_\mu \tan\beta & 
  4\mu v_d \Bigl( \lambda - \kappa \tan\beta 
    - \frac{\lambda A_\lambda \tan\beta}{2\mu} \Bigr)
\\
  4\mu v_u \Bigl( \lambda - \frac{\kappa}{\tan\beta} 
    - \frac{\lambda A_\lambda}{2\mu \tan\beta} \Bigr) & 
  4\mu v_d \Bigl( \lambda - \kappa \tan\beta 
    - \frac{\lambda A_\lambda \tan\beta}{2\mu} \Bigr) & 
  \frac{8\kappa B_\mu}{\lambda} \Bigl( 1 
    - \frac{\mu \left(A_\lambda + \frac{A_\kappa}{4}\right)}{B_\mu} 
    + \frac{\lambda^3 A_\lambda v_u v_d}{4 \kappa \mu B_\mu} \Bigr) 
  \end{matrix} \right),
\end{array}
\end{equation}
where $\bar{g}^2 \equiv g^2 + g'^2$, and we have assumed that all 
three expectation values are real and nonzero.  For us, the $A_\lambda$ 
and $A_\kappa$ terms are small because they are generated essentially 
only through weak renormalization group evolution below $k'$; 
$|A_{\lambda,\kappa}| \simlt O(10~{\rm GeV})$.  Other than contributing 
to $B_\mu$, they also contribute to singlet-doublet mixing and 
pseudoscalar masses, but we will ignore them in the following 
discussion on the (non-pseudo)scalar spectrum, as the result is 
not very sensitive to the values of such small $A$ terms.

\begin{figure}[t]
  \subfigure[$\tan\beta = 1.2$, $\kappa=1.0$, $\mu = 200~{\rm GeV}$]{\includegraphics[width=8.5cm]{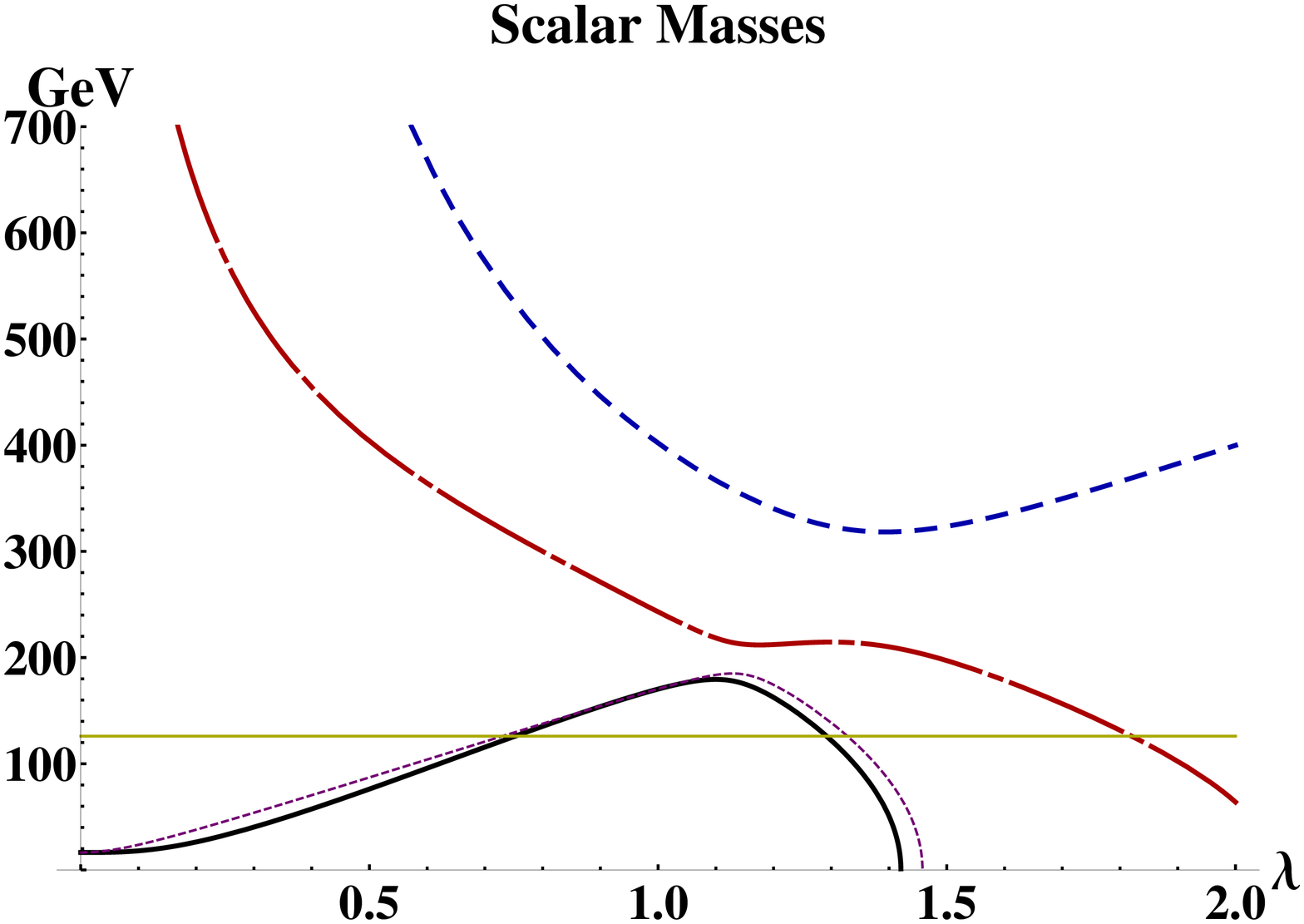}}
  \subfigure[$\tan\beta = 2.0$, $\kappa=1.0$, $\mu = 200~{\rm GeV}$]{\includegraphics[width=8.5cm]{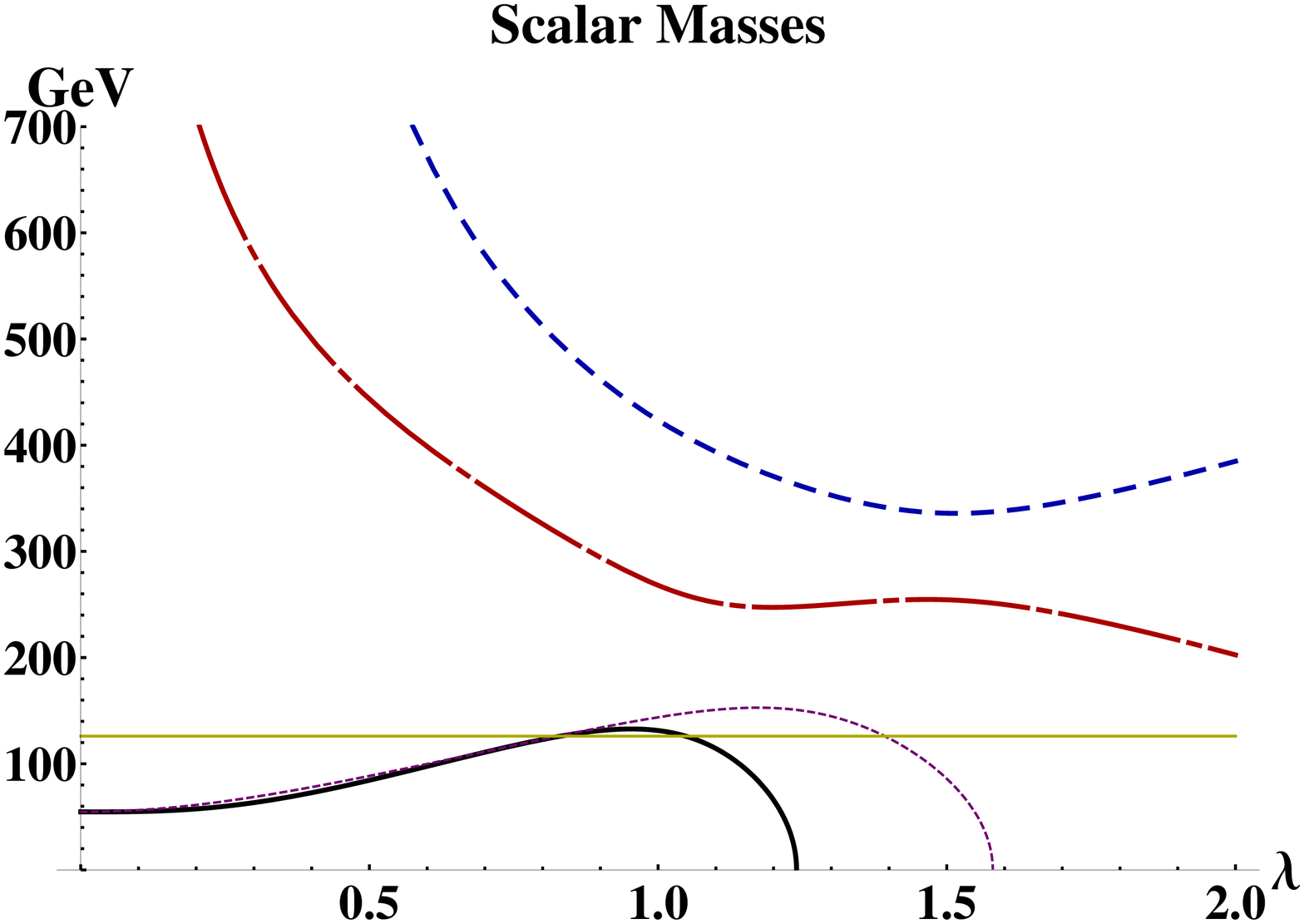}}
\caption{Two representative plots of the scalar mass spectrum in $\kappa$SUSY. 
 The solid (black), dot-dashed (red), and dashed (blue) lines represent 
 the masses of the three mass eigenstates, which at small $\lambda$ 
 correspond to the SM, heavy-doublet, and singlet like Higgs bosons, 
 respectively.  The horizontal (yellow) line shows $m_h = 125~{\rm GeV}$, 
 and the dotted (violet) line is the mass of the lightest Higgs boson 
 with singlet-doublet mixing turned off by hand.  In the left figure 
 we see that $\lambda$-doublet mixing is responsible for lowering the 
 mass of the Higgs below its decoupling limit, Eq.~(\ref{eq:decoup}), 
 rather than doublet-singlet mixing.  This is a generic feature for 
 $\tan\beta \sim 1$.  In the right figure, we see that as we increase 
 $\tan\beta$, singlet-doublet mixing sets in at lower $\lambda$ than 
 doublet-doublet mixing but that both are important in lowering the 
 Higgs mass below Eq.~(\ref{eq:decoup}).}
\label{fig:scalar}
\end{figure}

We now discuss important differences between $\kappa$SUSY 
and the MSSM as well as previous $\lambda$SUSY/NMSSM 
studies~\cite{Hall:2011aa,deGouvea:1997cx,Ellwanger:2009dp}. 
They are illustrated in Fig.~\ref{fig:scalar}, where (tree-level) 
scalar masses are plotted as a function of $\lambda$ for sample 
values of $\tan\beta, \kappa, \mu$.
\begin{itemize}
\item
We see that $\kappa$ plays a crucial role in this theory because it 
appears in $B_\mu \supset \kappa \mu^2/\lambda$.  It determines the 
degree of decoupling of the SM-like Higgs from the heavier scalars. 
The limit $\kappa = 0$ leads to nearly massless modes and is therefore 
unacceptable.  In fact, as we shall see, we need $\kappa \sim \lambda$ 
for a successful theory of electroweak symmetry breaking.
\item
The new quartic term $\lambda^2 |H_u H_d|^2$ leads to an extra 
doublet-doublet mixing which competes against $B_\mu$: ${\cal M}^2_{12} 
= (2\lambda^2 - \bar{g}^2/2) v_u v_d - B_\mu$.  As long as $2\lambda^2 
v_u v_d < B_\mu + \bar{g}^2 v_u v_d/2$, this leads to the well-known 
enhancement of the Higgs mass in $\lambda$SUSY, see Eq.~(\ref{eq:decoup}). 
However, once $2\lambda^2 v_u v_d > B_\mu + \bar{g}^2 v_u v_d/2$, the 
absolute magnitude of the off-diagonal term now increases with $\lambda$ 
which leads to lowering of the Higgs mass through the very same term. 
We call this effect $\lambda$-doublet mixing.  We find that in $\kappa$SUSY, 
this is the main effect that lowers the Higgs mass at large $\lambda$ 
and small $\tan\beta$, rather than mixing with the singlet, see 
Fig.~\ref{fig:scalar}.  This is different from Ref.~\cite{Hall:2011aa}, 
whose potential contains multiple extra free scales ($B_\mu$, the singlet 
mass) which are potentially large.  (Their benchmark point has $B_\mu 
\approx 4\mu^2 = (400~{\rm GeV})^2$.  In this region, $\lambda$-doublet 
mixing accounts for only $15\%$ of the lowering of the Higgs mass 
below its decoupling limit, Eq.~(\ref{eq:decoup}); the rest comes 
from singlet-doublet mixing.)  In fact, in $\kappa$SUSY, $\lambda 
\sim 2$ is excluded for $\mu \sim 200~{\rm GeV}$ exactly for this 
reason: the Higgs becomes tachyonic (i.e.\ the correct electroweak 
symmetry breaking vacuum disappears).
\item
The mass of the singlet-like scalar is not really a free parameter. 
It decouples together with the heavy Higgs ($B_\mu \rightarrow \infty$) 
but not independently.  This kind of relation is to be expected in 
a model with a scale-free superpotential, with $\lambda \sim \kappa$. 
It is simply an accidental feature (due to the coefficient in 
${\cal M}^2_{33}$) that the singlet-like scalar is heavier than 
other scalars by a factor of a few, in the limit of no mixing 
and $\lambda \sim \kappa$.
\item
Doublet-singlet mixing now depends on a difference between $\lambda$ 
and $\kappa$.  We find that, although the singlet-like Higgs is not 
very heavy, this greatly reduces mixing of the Higgs doublet component 
with the singlet and can lead to decoupling of the SM-like Higgs from 
the singlet for singlet masses as low as $400~{\rm GeV}$ for $\tan\beta 
\sim 1$.
\end{itemize}

In the limit of small $\lambda$-doublet mixing ($2\lambda^2 v_u v_d 
< B_\mu + \bar{g}^2 v_u v_d/2$) and negligible doublet-singlet mixing, 
the tree-level mass of the SM-like Higgs boson is given by
\begin{equation}
  m_h^2 \approx M_Z^2 \cos^2\!2\beta + \lambda^2 v^2 \sin^2\!2\beta.
\label{eq:decoup}
\end{equation}
In the appendix, we present analytical formulae for $m_h$ that include 
both large $\lambda$-doublet and singlet-doublet mixings up to second 
order in an expansion in $m_h^2/m_H^2$ (light Higgs mass over heavy 
Higgs mass); we also present the exact solution to $m_h$ in the regime 
where $\lambda$-doublet mixing dominates over doublet-singlet mixing 
as well as in the opposite case.

\begin{figure}[t]
  \subfigure[Higgs mass contours, $\kappa=0.7$, $\mu = 200~{\rm GeV}$]{\includegraphics[width=8cm]{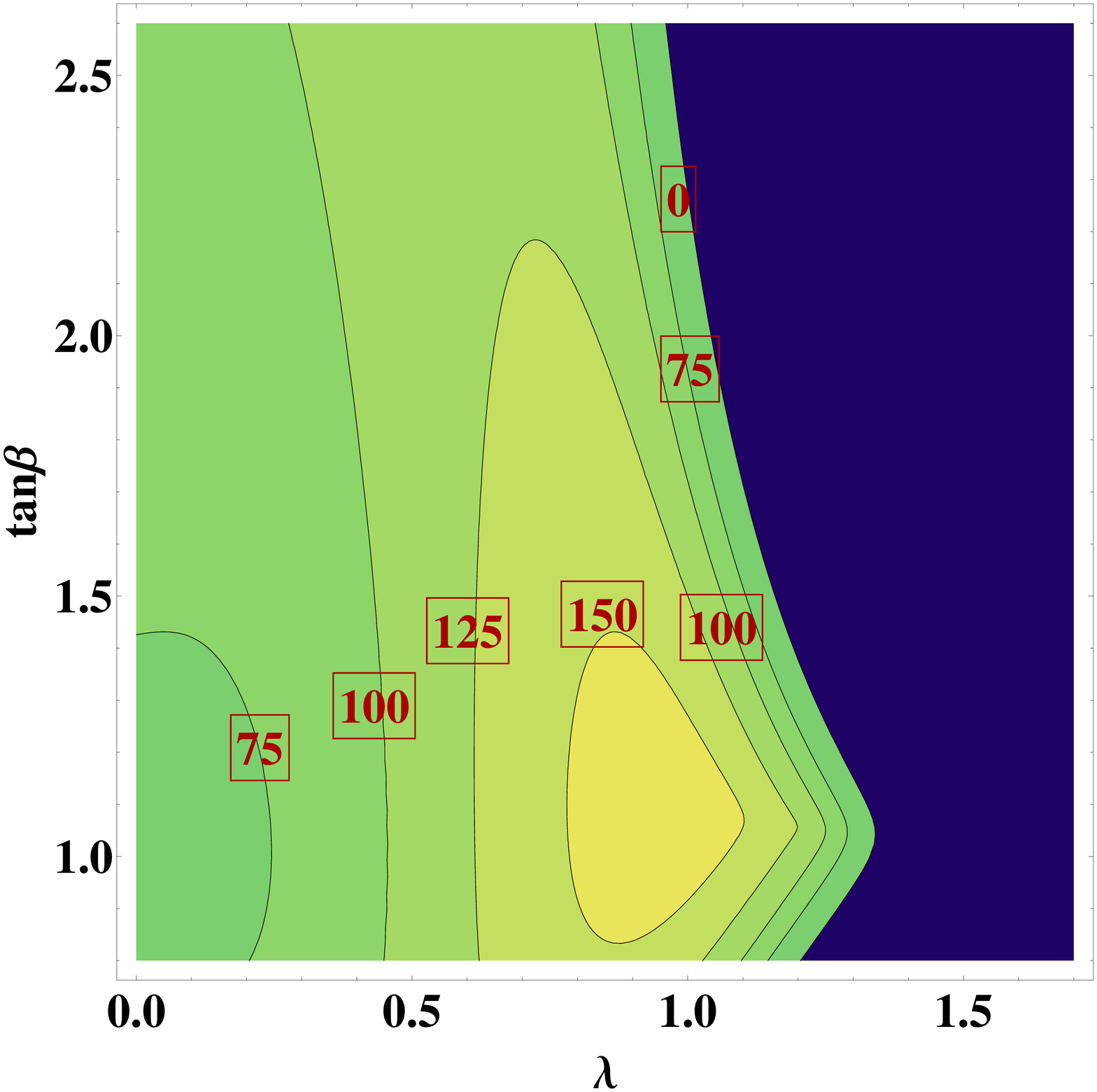}}
\hspace{0.5cm}
  \subfigure[Charged Higgs mass contours, $\kappa=0.7$, $\mu = 200~{\rm GeV}$]{\includegraphics[width=8cm]{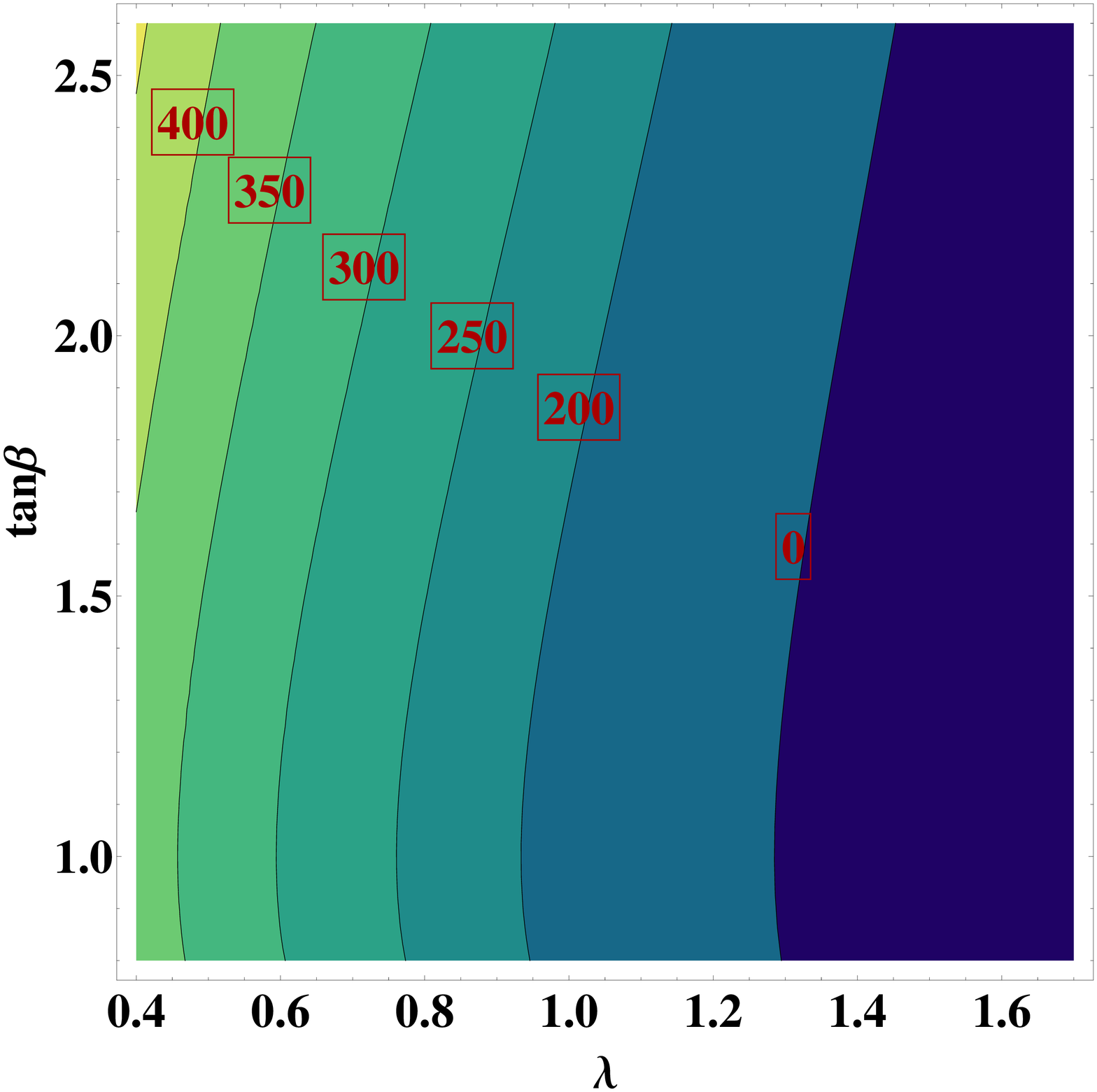}}
\caption{Left: Contours of the lightest Higgs boson mass in 
 $\lambda$-$\tan\beta$ plane.  We find the expected rise of the Higgs 
 mass with increasing $\lambda$ as well as the preference for low 
 $\tan\beta$.  The $\lambda$-doublet mixing effect is apparent for 
 large $\lambda$ where the Higgs mass quickly drops to zero.
\newline
 Right: Contours of the charged Higgs boson mass in the same plane. 
 Relatively low values of $\lambda$ are forced by the constraint 
 from $b \rightarrow s\gamma$, which requires $m_{H^+}$ not much 
 below $\sim 300~{\rm GeV}$.}
\label{fig:cont}
\end{figure}

In Fig.~\ref{fig:cont}(a), we present contours of the lightest Higgs boson 
mass, $m_h$, for typical values of parameters, where we have added the 
one-loop top-stop contribution with $m_{\tilde{t}} = 450~{\rm GeV}$ and 
$A_t = 0$.  In Fig.~\ref{fig:cont}(b), we show contours of the charged 
Higgs boson mass, which is given by
\begin{equation}
  m_{H^+}^2 = \frac{2 B_\mu}{\sin 2\beta} - \lambda^2 v^2 + M_W^2.
\label{eq:m_H+}
\end{equation}
In the non-decoupling region ($B_\mu/v^2 \simgt 1$) and for $\lambda 
> \sqrt{2}/\sin 2\beta$, the charged Higgs boson can become tachyonic. 
On the other hand, its mass cannot significantly fall below $300~{\rm GeV}$ 
due to constraints from $b \rightarrow s \gamma$.  This provides an 
important constraint on our parameter space and forces us to choose 
relatively low values of $\lambda \simlt 1$.

Another potential issue is a light pseudoscalar arising from an 
approximate $R$ symmetry under which $S, H_u, H_d$ have a charge of 
$2/3$.  This symmetry is spontaneously broken by $v_s, v_u, v_d$ so 
that there is a light $R$-axion.  This axion obtains a mass through 
loops of gauginos, mixing with other axions, such as the QCD axion, 
and $A_\lambda$, $A_\kappa$.  In Ref.~\cite{Dobrescu:2000yn}, it was 
determined that the $A$ terms provide the dominant contribution for 
$10^{-3} \simlt |A_{\lambda,\kappa}|/v \ll 1$, which we satisfy.  The 
mass of the $R$-axion due to the $A$ terms is given in terms of an 
expansion in $A_{\lambda,\kappa}/v$ by
\begin{equation}
  m_{A_1}^2 \approx 9 \frac{\mu}{\lambda} 
    \left( \frac{\lambda A_\lambda}{2} \frac{\cos^2\!\theta_A}{\sin 2\beta} 
    + \frac{\kappa A_\kappa}{3} \sin^2\!\theta_A \right) 
    + O(A_{\lambda,\kappa}^2),
\label{eq:R-axion}
\end{equation}
where $\tan\theta_A = \mu / (\lambda v \sin 2\beta) + O(A_{\lambda,\kappa}/v)$. 
We see that the mass is the geometric mean of $\mu$ and $A_{\lambda,\kappa}$ 
times $O(1)$ factors.  Since we generically have $|A_\lambda| > 1~{\rm GeV}$, 
the mass is in tens of GeVs, so we are safe from the constraint from 
$\Upsilon$ decays.  Since $\lambda,\kappa$ are $O(1)$, however, the 
Higgs can also decay into the $R$-axion with a large branching fraction, 
if this decay mode is kinematically allowed.  Assuming $m_h = 125~{\rm GeV}$, 
we find that this happens for $|A_\lambda| < 10~{\rm GeV}$ (neglecting 
$A_\kappa$).  Depending on parameters, we can have $|A_\lambda| \simgt 
10~{\rm GeV}$, in which case decays of the lightest Higgs boson are 
SM-like.

\subsection{Sample spectra}
\label{subsec:spectra}

We here present sample parameter points in $\kappa$SUSY.  To achieve 
successful electroweak symmetry breaking, in particular to obtain a 
sufficiently large $\mu = \lambda v_s$, we introduce a negative soft 
mass-squared for the singlet at $k'$, $m_S^2 \sim -(400~{\rm GeV})^2$. 
Such a term can arise naturally if there are (additional) messenger 
fields $f, \bar{f}$ on the IR brane which couple to the $S$ field 
in the superpotential $W = S f \bar{f}$~\cite{Dine:1996xk}.  Here 
$f, \bar{f}$ are assumed to be SM-gauge singlets and have supersymmetric 
and supersymmetry breaking masses (roughly) of order $k'$: $M_f \sim 
\sqrt{F}_f \sim k'$.  (This does not require a strong coincidence because 
the characteristic scale on the IR brane is $k' \sim M_*'$.)  The $A$ 
terms generated by $f,\bar{f}$ loops are small for $M_f \sim \sqrt{F}_f$, 
since both the $A$ terms and the soft mass-{\it squared}, $m_S^2$, 
are generated at the one-loop order.

We present two sample spectra in Figs.~\ref{fig:spec1} and \ref{fig:spec2}, 
which correspond respectively to two different choices of the compositeness 
scale, $k' = 10~{\rm TeV}$ and $1000~{\rm TeV}$, and will be discussed in 
more detail in Sections~\ref{subsubsec:k'=10} and \ref{subsubsec:k'=1000}. 
The relevant parameters for electroweak symmetry breaking are $\lambda$, 
$\kappa$, $m_S^2$, and the electroweak gaugino masses $M_{1,2}$.  (We 
choose $m_{H_u,H_d,\tilde{Q}_3,\tilde{U}_3}^2, A_{\lambda,\kappa,t,b} 
\approx 0$ at $k'$, ignoring loop-suppressed threshold corrections.) 
The gluino mass is chosen to be small (but still allowed by the experimental 
constraint) to alleviate fine-tuning, and the bino is chosen to be the 
lightest observable-sector supersymmetric particle (LOSP).  For the gluino 
mass we add the one-loop threshold correction, which can be as large as 
$\approx 20\%$ for the multi-TeV squark masses~\cite{Martin:2005ch}.  In 
this section, we assume that the gravitino is heavier than the LOSP, so 
that the bino is the lightest supersymmetric particle.  This is the case 
for $M_{\rm mess} \simgt M_{\rm Pl}$, or for $M_{\rm mess} \simlt M_{\rm Pl}$ 
if there is additional supersymmetry breaking that does not contribute 
to the MSSM superparticle masses but pushes up the gravitino mass 
above the LOSP mass~\cite{Nomura:2001ub}.  If the gravitino is lighter 
than the bino, somewhat stronger bounds on the gluino mass would 
apply~\cite{Kats:2011qh}.  For example, if the bino decays promptly 
to the gravitino, then the lower bound is $m_{\tilde{g}} \approx 
900~{\rm GeV}$, instead of $\approx 700~{\rm GeV}$.

In presenting the sample points, we also evaluate the amount of fine-tuning, 
adopting a conventional criterion~\cite{Barbieri:1987fn}
\begin{equation}
  \Delta = \max_{i,j} \frac{d\ln A_i}{d\ln B_j},
\label{eq:def-Delta}
\end{equation}
where $A_i = (m_h^2, v^2)$ and $B_j$ are UV parameters to be specified 
below.  The $A_i$ correspond to the $(\theta_{h,h_u}\hat v_u + 
\theta_{h,h_d} \hat v_d + \theta_{h,s} \hat v_s)$ and $(\vec v_u 
+ \vec v_d)/v$ directions in the three-dimensional $v_u,v_d,v_s$ space, 
respectively, where we define scalar mixing angles in terms of eigenvector 
overlap: $h = \theta_{h,h_u} h_u + \theta_{h,h_d} h_d + \theta_{h,s} s$. 
In the case of $\lambda$-doublet or singlet-doublet mixing, fine-tuning 
(e.g.\ due to stops) may be much alleviated compared to the MSSM due to 
level repulsion which is generated naturally through large off-diagonal 
elements in the mass matrix; in the case of singlet-doublet mixing, 
this has been analyzed in Ref.~\cite{Hall:2011aa}.  We here point out 
that large-mixing, natural scenarios with TeV-scale stops are typically 
accompanied by drastic deviations of Higgs couplings, so if the Higgs 
has only moderate deviations from SM cross sections and decay rates, 
then naturalness generically requires light stops.  In our analysis, 
we choose $B_j = (\lambda, \kappa, m_S^2, M_{1,2,3}, k', y_t, \tilde{m})$.

\subsubsection{\boldmath $k'=10~{\rm TeV}$}
\label{subsubsec:k'=10}

The following considerations give a bottom-up picture of what is 
needed to generate a natural superpartner spectrum (in the decoupling 
regime)~\cite{Kitano:2006gv,Papucci:2011wy} that radiatively breaks 
electroweak symmetry with $k' = 10~{\rm TeV}$:
\begin{itemize}
\item The fine-tuning constraint ($\Delta^{-1} \simgt 20\%$) requires 
$|\mu| \simlt 210~{\rm GeV}$, $m_{\tilde{t}} \simlt 410~{\rm GeV}$ 
(for degenerate stop masses without mixing), $m_{\tilde{g}} \simlt 
790~{\rm GeV}$ (at the leading-log level; the actual bound is significantly 
weaker because of the effect of strong interactions), $m_{\tilde{W}} 
\simlt 890~{\rm GeV}$, $m_{\tilde{B}} \simlt 2800~{\rm GeV}$, and 
$\tilde{m} \simlt 4~{\rm TeV}$.
\item Electroweak symmetry breaking requires $\lambda, \kappa \sim 0.7$ at 
low energies, as discussed in the last section; we also need $m_S^2 \sim 
-(400~{\rm GeV})^2$ to generate a sufficiently large $\mu$ term.
\end{itemize}
\begin{figure}[t]
\begin{minipage}{2.4in}
\begin{picture}(330,300)(-50,-45)
  \LongArrow(0,-20)(0,225) \Text(-5,225)[br]{$E$}
  \Line(-3,205)(3,205) \Text(-8,205)[r]{\small $10~{\rm TeV}$}
  \Line(-3,175)(3,175) \Text(-8,177)[r]{\small $4~{\rm TeV}$}
                       \Text(-8,155)[r]{\small \vdots}
  \Line(-3,120)(3,120) \Text(-8,122)[r]{\small $800~{\rm GeV}$}
  \Line(-3,85)(3,85) \Text(-8,87)[r]{\small $600~{\rm GeV}$}
  \Line(-3,50)(3,50) \Text(-8,52)[r]{\small $400~{\rm GeV}$}
  \Line(-3,15)(3,15) \Text(-8,17)[r]{\small $200~{\rm GeV}$}
  \Line(-3,-20)(3,-20) \Text(-8,-18)[r]{\small $0$}
  \Line(25,205)(95,205) \Text(60,209)[b]{KK states}
  \Line(25,175)(95,175) \Text(60,177)[b]{$\Psi_{1,2}$ ($D_3, L_3, E_3$)}
  \Line(25,140)(95,140) \Text(60,142)[b]{$\tilde{g}$}
  \Line(25,106)(95,106) \Text(60,108)[b]{$\tilde{W}^+, \tilde{W}^0$}
  \Line(25,103)(95,103)
  \Line(25,59)(95,59) \Text(60,61)[b]{$\tilde{t}_2, \tilde{t}_1, \tilde{b}_1$}
  \Line(25,55)(95,55)
  \Line(25,53)(95,53)
  \Line(25,18)(95,18) \Text(60,20)[b]{$\tilde{h}^0_2, \tilde{h}^+,\tilde{h}^0_1, \tilde{B}$}
  \Line(25,15)(95,15)
  \Line(25,8)(95,8)
  \Line(25,3)(95,3)
\end{picture}
\end{minipage}
\begin{minipage}{3in}
\frame{
\begin{tabular}{c|c|c}
  Parameters & Spectrum (GeV) & Properties \\ \hline \hline
  $\lambda = 0.66$ & $m_h = 125$ & $\mu = -201~{\rm GeV}$ \\
  $\kappa = 0.69$  & $m_H = 322$ & $\tan\beta = 1.97$ \\
  $m_S^2 = -(328~{\rm GeV})^2$ & $m_s = 436$ & $\theta_{h,s} = 0.03$ \\
  $k' = 10~{\rm TeV}$ & $m_{A_s} = 71$ & $\theta_{H,s} = 0.38$ \\
  $\tilde{m} = 4~{\rm TeV}$ & $m_A = 367$ & $A_\lambda = -21~{\rm GeV}$ \\
  & $m_{H^+} = 329$ & $A_\kappa = -2~{\rm GeV}$ \\
  & $m_{\tilde{g}} = 946$ & $A_t = 217~{\rm GeV}$ \\
  & $m_{\tilde{W}^+,\tilde{W}^0} = 708$ & \\
  & $m_{\tilde{B}} = 127$ & \\
  & $m_{\tilde{t}_1} = 431$ & \\
  & $m_{\tilde{t}_2} = 467$ & \\
  & $m_{\tilde{b}_1} = 416$ & \\
  & $m_{\tilde{h}^0_2} = 214$ &\\
  & $m_{\tilde{h}^+} = 206$ &\\
  & $m_{\tilde{h}^0_1} = 162$ &\\
  & $m_{\tilde{s}} = 467$ &\\
\end{tabular}
}
\end{minipage}
\caption{A typical mass spectrum for a compositeness scale of $k' = 
 10~{\rm TeV}$.  The states with mixing are labeled by their largest 
 components.  In the left diagram, the states are always ordered from 
 heavy to light.  The gluino mass of $m_{\tilde{g}} = 946~{\rm GeV}$ 
 in the table corresponds to $M_3 = 801~{\rm GeV}$ at the scale 
 $m_{\tilde{t}}$ obtained using the MSSM renormalization group evolution. 
 The wino is relatively heavy, which is necessary to generate a mass 
 for the light pseudoscalar $m_{A_s} > m_h/2$ through the $A_\lambda$ 
 term, in line with recent hints of a Higgs discovery.  If the wino is 
 much lighter, the Higgs would decay almost entirely to pseudoscalars.}
\label{fig:spec1}
\end{figure}
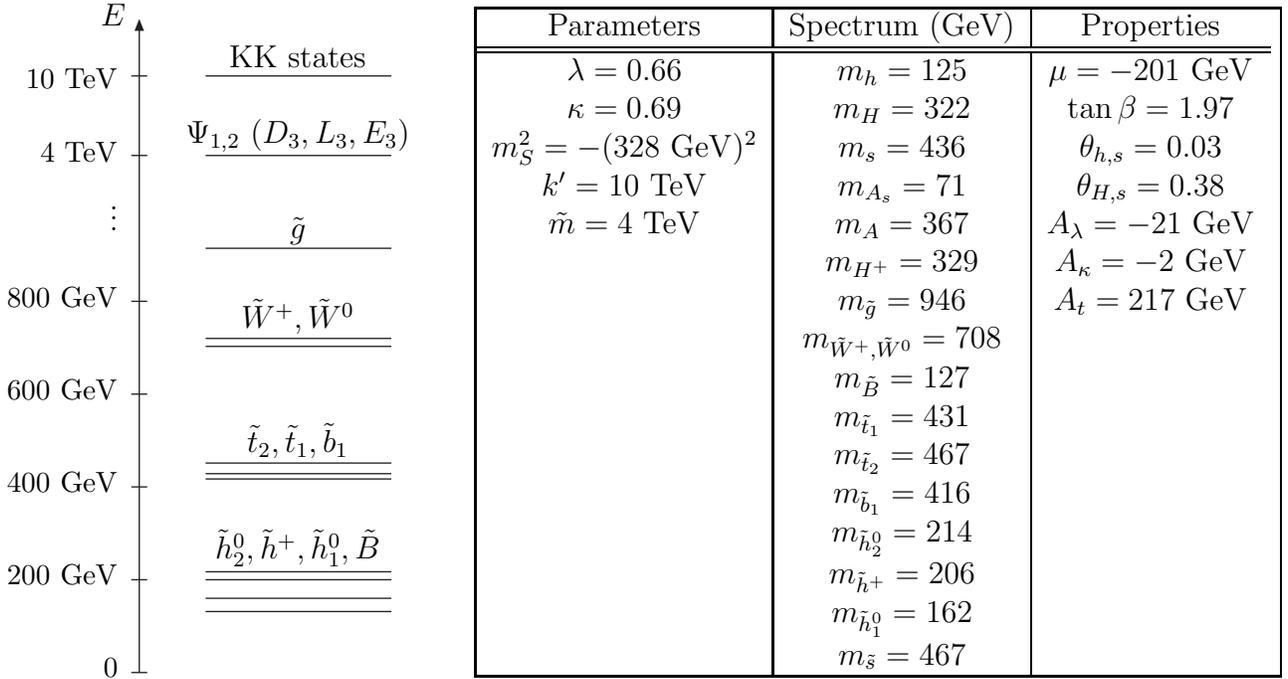

In Fig.~\ref{fig:spec1}, we show a typical mass spectrum for $k' = 
10~{\rm TeV}$, where the lightest Higgs boson mass is evaluated with 
the one-loop top-stop contribution added.  The production cross section 
$\sigma(g g \rightarrow h)$ is modified relative to the SM due to 
non-decoupling stop contributions and $A$ terms; this sample point 
has an enhancement of $13\%$.  Unlike in the MSSM, the decay rate 
of the Higgs into $\bar{b}b$ is depleted in $\lambda$SUSY relative 
to the SM rate.  As expanded in $m_h^2/m_H^2$, the rate is given by 
(see the appendix)
\begin{equation}
  \frac{\Gamma(h \rightarrow \bar{b}b)}{\Gamma_{\rm SM}(h \rightarrow 
    \bar{b}b)} = 1 - \tan\beta\, \sin^2\! 2\beta\, 
    \frac{\lambda v \sqrt{|\lambda^2 v^2 - M_Z^2|}} {2 B_\mu}.
\label{eq:h_to_bb}
\end{equation}
For the $k'=10~{\rm TeV}$ spectrum, this formula gives $0.88$, within 
$10\%$ of the exact result, $0.96$.  Because of this suppression, the 
branching ratios into other modes are enhanced.  In particular, we find 
that ${\rm Br}(h \rightarrow \gamma\gamma)$ is increased by $4\%$ with 
respect to the SM, resulting in an enhancement of $\sigma(g g \rightarrow h) 
\times {\rm Br}(h \rightarrow \gamma\gamma)$ of $18\%$.  This effect 
of an enhanced $\gamma\gamma$ signal has been observed for a different 
parameter space of $\lambda$SUSY in Ref.~\cite{Hall:2011aa}; however, 
here the effect is not large and the decays are practically SM-like. 
Notice in particular the small mixing of the Higgs with the singlet as 
anticipated in section~\ref{subsec:Higgs}.  For our $k' = 10~{\rm TeV}$ 
point, decay rate times branching ratio of both the heavy Higgs and the 
singlet into $WW$ or $ZZ$ is four orders of magnitude below that of the 
SM Higgs of the same mass, which makes them invisible to SM Higgs searches. 
For the fine-tuning parameter, we obtain $\Delta^{-1} = 19\%$, consistent 
with expectations based on the general argument.

The heavy Higgs decays into $A_s A_s$, $A_s Z$, and $\bar{t}t$, with 
$A_s$ decaying predominantly into $\bar{b}b$.  The singlet decays into 
$A_s A_s$, $A_s Z$, $\bar{t}t$, and $\tilde{h}^+ \tilde{h}^-$.  Due to 
associated $Z$ production, discovery of these particles may be possible 
at $e^+ e^-$ colliders such as the ILC/CLIC.

\subsubsection{\boldmath $k'=1000~{\rm TeV}$}
\label{subsubsec:k'=1000}

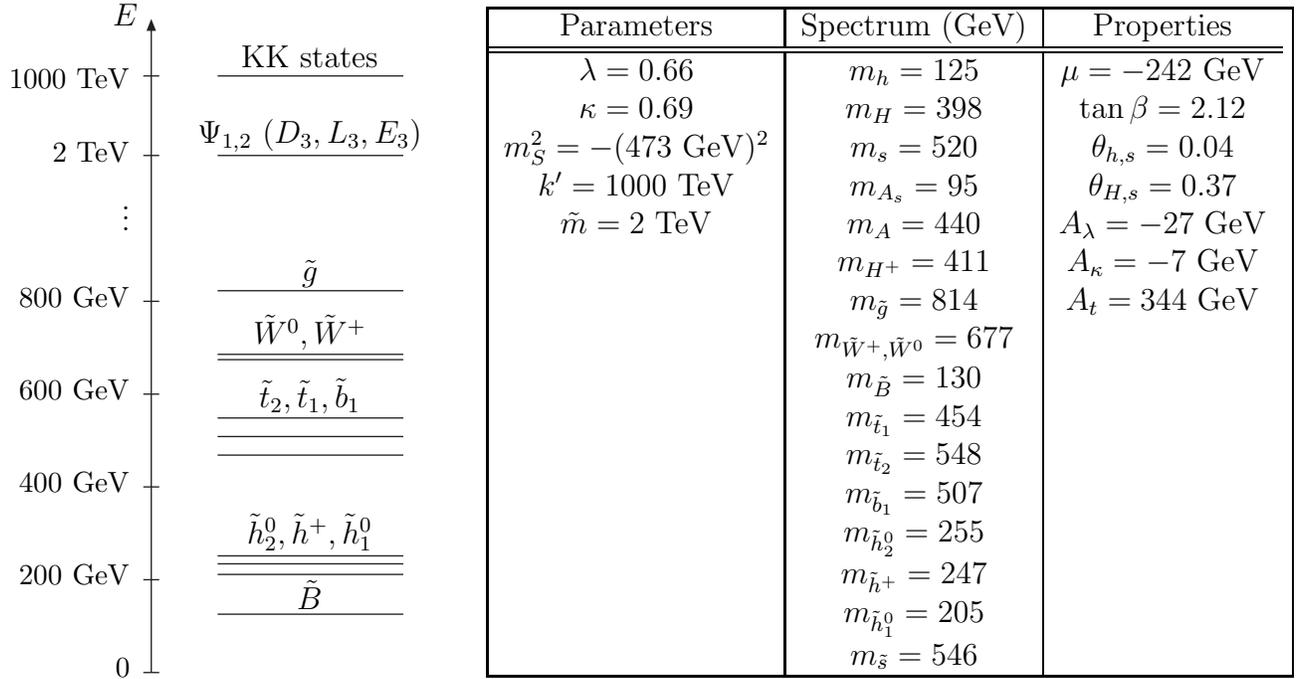
\begin{figure}[t]
\begin{minipage}{2.4in}
\begin{picture}(330,300)(-50,-45)
  \LongArrow(0,-20)(0,225) \Text(-5,225)[br]{$E$}
  \Line(-3,205)(3,205) \Text(-8,205)[r]{\small $1000~{\rm TeV}$}
  \Line(-3,175)(3,175) \Text(-8,177)[r]{\small $2~{\rm TeV}$}
                       \Text(-8,155)[r]{\small \vdots}
  \Line(-3,120)(3,120) \Text(-8,122)[r]{\small $800~{\rm GeV}$}
  \Line(-3,85)(3,85) \Text(-8,87)[r]{\small $600~{\rm GeV}$}
  \Line(-3,50)(3,50) \Text(-8,52)[r]{\small $400~{\rm GeV}$}
  \Line(-3,15)(3,15) \Text(-8,17)[r]{\small $200~{\rm GeV}$}
  \Line(-3,-20)(3,-20) \Text(-8,-18)[r]{\small $0$}
  \Line(25,205)(95,205)  \Text(60,209)[b]{KK states}
  \Line(25,175)(95,175) \Text(60,177)[b]{$\Psi_{1,2}$ ($D_3, L_3, E_3$)}
  \Line(25,124)(95,124) \Text(60,126)[b]{$\tilde{g}$}
  \Line(25,100)(95,100) \Text(60,102)[b]{$\tilde{W}^0, \tilde{W}^+$}
  \Line(25,98)(95,98)
  \Line(25,76)(95,76) \Text(60,78)[b]{$\tilde{t}_2, \tilde{t}_1, \tilde{b}_1$}
  \Line(25,69)(95,69)
  \Line(25,62)(95,62)
  \Line(25,24)(95,24) \Text(60,26)[b]{$\tilde{h}^0_2, \tilde{h}^+, \tilde{h}^0_1$}
  \Line(25,21)(95,21)
  \Line(25,17)(95,17)
  \Line(25,02)(95,02) \Text(60,04)[b]{$\tilde{B}$}
\end{picture}
\end{minipage}
\begin{minipage}{3in}
\frame{
\begin{tabular}{c|c|c}
Parameters & Spectrum (GeV) & Properties \\ \hline \hline
  $\lambda = 0.66$ & $m_h = 125$ & $\mu = -242~{\rm GeV}$ \\
  $\kappa = 0.69$  & $m_H = 398$ & $\tan\beta = 2.12$ \\
  $m_S^2 = -(473~{\rm GeV})^2$ & $m_s = 520$ & $\theta_{h,s} = 0.04$ \\
  $k' = 1000~{\rm TeV}$ & $m_{A_s} = 95$ & $\theta_{H,s} = 0.37$ \\
  $\tilde{m} = 2~{\rm TeV}$ & $m_A = 440$ & $A_\lambda = -27~{\rm GeV}$ \\
  & $m_{H^+} = 411$ & $A_\kappa = -7~{\rm GeV}$ \\
  & $m_{\tilde{g}} = 814$ & $A_t = 344~{\rm GeV}$ \\
  & $m_{\tilde{W}^+,\tilde{W}^0} = 677$ & \\
  & $m_{\tilde{B}} = 130$ & \\
  & $m_{\tilde{t}_1} = 454$ & \\
  & $m_{\tilde{t}_2} = 548$ & \\
  & $m_{\tilde{b}_1} = 507$ & \\
  & $m_{\tilde{h}^0_2} = 255$ &\\
  & $m_{\tilde{h}^+} = 247$ &\\
  & $m_{\tilde{h}^0_1} = 205$ &\\
  & $m_{\tilde{s}} = 546$ &\\
\end{tabular}
}
\end{minipage}
\caption{A typical mass spectrum for a compositeness scale of $k' = 
 1000~{\rm TeV}$.  Definitions are as in Fig.~\ref{fig:spec1}.  The 
 gluino mass of $m_{\tilde{g}} = 814~{\rm GeV}$ corresponds to $M_3 
 = 710~{\rm GeV}$ at the scale $m_{\tilde{t}}$ obtained using the MSSM 
 renormalization group evolution.}
\label{fig:spec2}
\end{figure}

The fine-tuning constraint will be more severe for $k' = 1000~{\rm TeV}$ 
than for $k' = 10~{\rm TeV}$ because of the large $\ln(k'/{\rm TeV}) = 6.9$. 
Performing the same bottom-up analysis as in the case of $k' = 10~{\rm TeV}$, 
we find:
\begin{itemize}
\item The fine-tuning constraint ($\Delta^{-1} \simgt 10\%$) requires 
$|\mu| \simlt 290~{\rm GeV}$, $m_{\tilde{t}} \simlt 370~{\rm GeV}$ (for 
degenerate stop masses without mixing), $m_{\tilde{g}} \simlt 460~{\rm GeV}$ 
(again at the leading-log level), $m_{\tilde{W}} \simlt 800~{\rm GeV}$, 
$m_{\tilde{B}} \simlt 2500~{\rm GeV}$, and $\tilde{m} \simlt 3.6~{\rm TeV}$.
\end{itemize}
For $k' = 1000~{\rm TeV}$, the theory is expected to be fine-tuned 
at the $10\%$ level.

In Fig.~\ref{fig:spec2}, we show a typical mass spectrum for $k' = 
1000~{\rm TeV}$.  We find that, as in the $k' = 10~{\rm TeV}$ case, the 
phenomenology of the Higgs is mostly SM-like: the production cross section 
$\sigma(g g \rightarrow h)$ is enhanced by $9\%$ relative to the SM; 
Eq.~(\ref{eq:h_to_bb}) gives $0.92$ as the decay rate of the Higgs to 
$\bar{b}b$ with respective to the SM, whereas the exact result is $0.96$. 
This translates into an increase of $\sigma(gg \rightarrow h) \times 
{\rm Br}(h \rightarrow \gamma\gamma)$ of $13\%$ with respect to the SM. 
As for the heavy Higgs or the singlet, we again find four orders of 
magnitude suppression of production cross section times branching 
ratio into $WW$ or $ZZ$ compared to the SM Higgs with the same mass. 
We find fine-tuning of $\Delta^{-1} = 10\%$ for this sample point, 
which is in agreement with expectations.

We find that if we relax our requirement of tuning slightly, we can 
choose $k'$ to be much larger than $1000~{\rm TeV}$ without conflicting 
with Landau pole constraints.  We, however, note that two-loop stop 
contributions to the Higgs quartic are negative and the theory will 
therefore require larger $\lambda, \kappa$, so it is not obvious that 
this statement will hold at two loops.  Using the tree-level potential 
as the other extreme to the one-loop potential, one finds that large 
$\lambda, \kappa \sim 0.8~\mbox{--}~0.9$ are needed to push the Higgs 
mass high enough and one cannot take $k'$ much higher than $1000~{\rm TeV}$ 
due to Landau pole constraints.  The truth is expected to lie somewhere 
between the tree-level and one-loop situations.

\section{Flat Space Realization}
\label{sec:flat}

We now discuss realizing our basic setup, Fig.~\ref{fig:setup}, using 
a flat space extra dimension.  An obvious way to do this is to simply 
turn off the curvature in models of Section~\ref{sec:model}.  The 
analysis then goes similarly with the replacement $k' \rightarrow 
1/\pi R$, except that we now do not have a large desert above the 
compactification scale, $1/R$, so we cannot have the high-scale see-saw 
mechanism or conventional gauge coupling unification.

In this section, we pursue an alternative realization, adopting 
supersymmetry breaking by boundary conditions associated with a compact 
extra dimension~\cite{Pomarol:1998sd}.  Our model is essentially that 
in Ref.~\cite{Barbieri:2001yz}.  Specifically, we consider an $SU(3)_C 
\otimes SU(2)_L \otimes U(1)_Y$ gauge theory in 5D, with the extra 
dimension compactified on $S^1/Z_2$: $0 \leq y \leq \pi R$.  We introduce 
three generations of matter and Higgs fields in the bulk, but localize 
the third-generation quark doublet, right-handed top, and Higgs 
multiplets to the $y=\pi R$ brane:
\begin{equation}
  M_{Q_3,U_3,H_u,H_d} \ll -\frac{1}{\pi R},
\label{eq:flat-M-3}
\end{equation}
where $M_{\Phi}$ represents bulk masses as in previous sections.  When 
supersymmetry is broken by twisted boundary conditions with twist parameter 
$\alpha$, we obtain
\begin{equation}
  m_{\tilde{Q}_3, \tilde{U}_3, H_u, H_d} \ll 
    m_{\tilde{\Psi}_{1,2}, \tilde{D}_3, \tilde{L}_3, \tilde{E}_3} 
  = \frac{\alpha}{R},
\label{eq:flat-soft}
\end{equation}
at the scale $1/R$, where we have taken
\begin{equation}
  |M_{\Psi_{1,2}, D_3, L_3, E_3}| \ll \frac{1}{\pi R}.
\label{eq:flat-M-12}
\end{equation}
This condition guarantees that the first two generation sfermions are 
nearly degenerate in mass, avoiding stringent constraints from flavor 
violation.

To obtain the spectrum we want, we take $\alpha/R$ to be in the multi-TeV 
region.  For the gauge multiplets, we introduce sizable gauge kinetic 
terms on (one or both of) branes, which control the size of the gaugino 
masses:
\begin{equation}
  M_A = \frac{\pi R g_{4,A}^2}{g_{5,A}^2} \frac{\alpha}{R},
\end{equation}
where $g_{5,A}$ and $g_{4,A}$ are the 5D bulk and 4D gauge couplings, 
respectively, with $g_{4,A}$ given by
\begin{equation}
  \frac{1}{g_{4,A}^2} = \frac{\pi R}{g_{5,A}^2} 
    + \frac{1}{\tilde{g}_{0,A}^2} + \frac{1}{\tilde{g}_{\pi,A}^2},
\label{eq:4D-5D-gauge}
\end{equation}
in terms of $g_{5,A}$ and the brane-localized gauge couplings at $y=0$ 
and $\pi R$, $\tilde{g}_{0,A}$ and $\tilde{g}_{\pi,A}$.  We take $M_A$ 
to be in the sub-TeV region.

Introducing a singlet field $S$ together with the superpotential $\lambda 
S H_u H_d + f(S)$ on the $y=\pi R$ brane, the analysis of electroweak 
symmetry breaking goes as in the previous section, with the identification
\begin{equation}
  k' \rightarrow \frac{1}{\pi R},
\qquad
  \tilde{m} \rightarrow \frac{\alpha}{R}.
\label{eq:flat-corresp}
\end{equation}
A negative soft mass-squared for $S$ can be induced, for example, by 
introducing some bulk field that has a Yukawa coupling to $S$ on the 
$y=\pi R$ brane.

In the present model, the two circles in Fig.~\ref{fig:setup} are 
interpreted as the 5D bulk (left) and the $y=\pi R$ brane (right), 
rather than the $y=0$ and $\pi R$ branes as in previous models. 
Because of Eq.~(\ref{eq:flat-M-12}), only a part of the Yukawa 
hierarchy can be explained by wavefunction profiles.  With 
Eqs.~(\ref{eq:flat-M-3},~\ref{eq:flat-M-12}) the Yukawa matrices 
obtain the following structure from the wavefunctions:
\begin{equation}
  y_u \sim 
  \left( \begin{array}{ccc}
    \epsilon^2 & \epsilon^2 & \epsilon \\
    \epsilon^2 & \epsilon^2 & \epsilon \\
    \epsilon & \epsilon & 1
  \end{array} \right),
\qquad
  y_d \sim
  \left( \begin{array}{ccc}
    \epsilon^2 & \epsilon^2 & \epsilon^2 \\
    \epsilon^2 & \epsilon^2 & \epsilon^2 \\
    \epsilon & \epsilon & \epsilon
  \end{array} \right),
\qquad
  y_e \sim
  \left( \begin{array}{ccc}
    \epsilon^2 & \epsilon^2 & \epsilon^2 \\
    \epsilon^2 & \epsilon^2 & \epsilon^2 \\
    \epsilon^2 & \epsilon^2 & \epsilon^2
  \end{array} \right),
\label{eq:flat-y}
\end{equation}
where $O(1)$ factors are omitted in each element, and $\epsilon 
\equiv 1/\sqrt{\pi R}$ is the volume dilution factor.  The structure 
beyond this must come from that of 5D Yukawa couplings between matter 
and Higgs on the $y=\pi R$ brane.

\section{Conclusions}
\label{sec:concl}

In this paper we have presented supersymmetric models in which light 
stops are obtained while keeping near flavor degeneracy for the first 
two generation sfermions.  Such a spectrum is motivated by the naturalness 
argument together with the recent LHC data.  Our construction is based 
on the basic picture in Fig.~\ref{fig:setup}: being ``close to'' electroweak 
symmetry breaking implies being ``away from'' supersymmetry breaking, 
and vice versa.  In models where the two sectors correspond to the 
two branes at the opposite ends of a (warped or flat) extra dimension, 
the desired superpartner spectra are obtained while reproducing the 
hierarchy in the Yukawa matrices through wavefunction profiles of 
the quark/lepton fields.  A relatively large Higgs boson mass, including 
$m_h = 125~{\rm GeV}$, can be easily accommodated if the scale of 
Kaluza-Klein excitations is low.  For models in warped space, the 
AdS/CFT correspondence allows us to interpret them in terms of purely 
4D theories in which the top and Higgs (and the left-handed bottom) 
multiplets are composites of some strongly interacting sector.  An 
alternative realization of the picture in Fig.~\ref{fig:setup} is 
obtained by identifying the two sectors as the bulk of a flat extra 
dimension and a brane on its boundary, and by breaking supersymmetry 
by boundary conditions, which we have also discussed.

In the coming years, the LHC will be exploring the parameter regions of 
supersymmetric theories in which the stops (and the left-handed sbottom) 
are light.  If electroweak symmetry breaking is indeed natural in the 
conventional sense, the LHC will find the stops in the sub-TeV region. 
If not, and if the SM-like Higgs boson is confirmed with $m_h \simeq 
125~{\rm GeV}$, then we would be led to consider that supersymmetry 
is absent at low energies, or it is realized in a somewhat 
fine-tuned form, perhaps along the lines of scenarios considered 
in Refs.~\cite{Wells:2003tf,ArkaniHamed:2004fb,Hall:2011jd}.

\section*{Acknowledgments}

We thank David Pinner, Joshua Ruderman, and Satoshi Shirai for useful 
discussions.  This work was supported in part by the Director, Office 
of Science, Office of High Energy and Nuclear Physics, of the US Department 
of Energy under Contract DE-AC02-05CH11231 and by the National Science 
Foundation under grants PHY-0855653.

\appendix

\section{Analytical Formulae for the Higgs Boson Mass}
\label{app:Higgs}

\begin{figure}[t]
  \subfigure[$\tan\beta = 1.2$, $\kappa=1.0$, $\mu = 200~{\rm GeV}$]{\includegraphics[width=9cm]{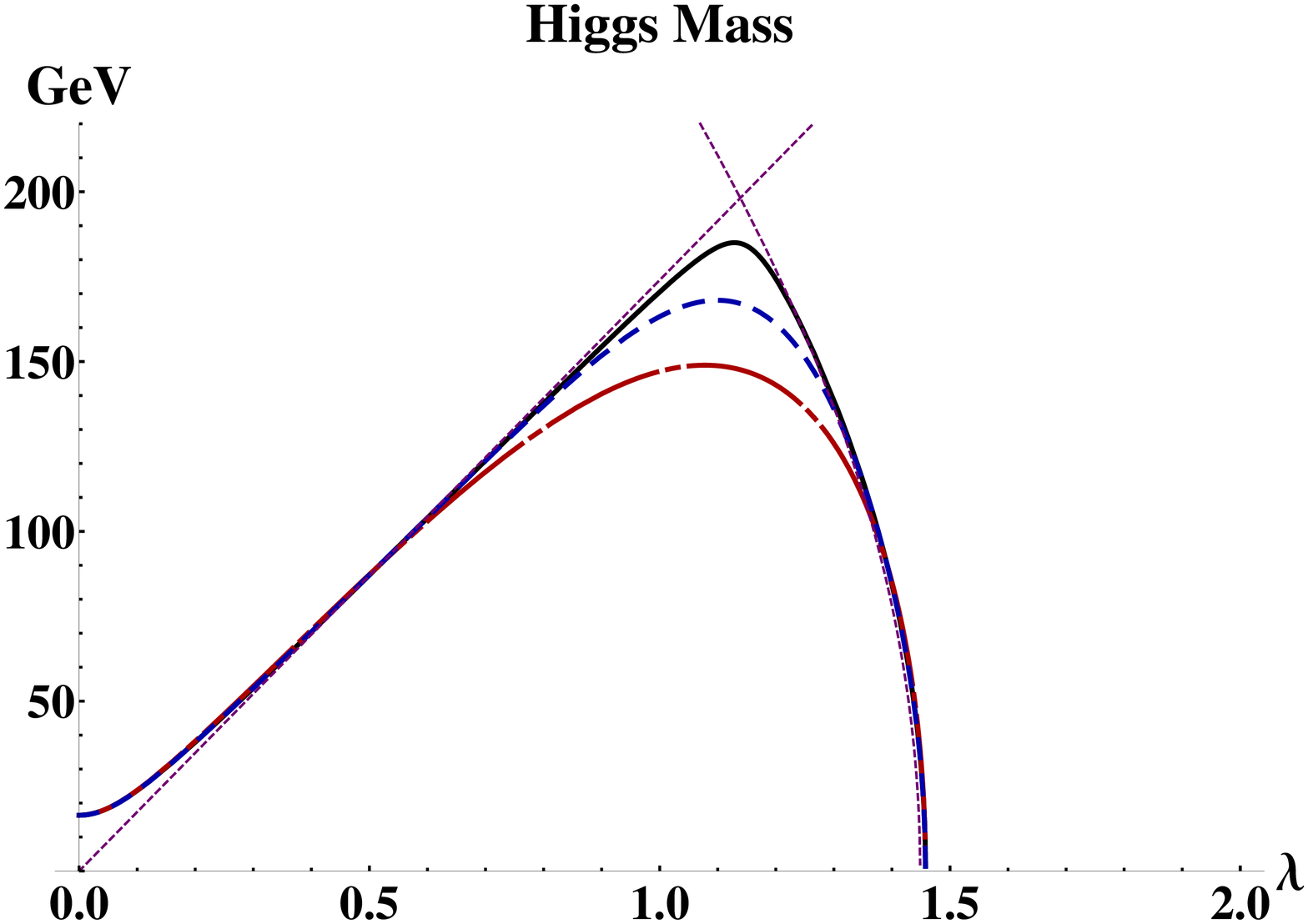}}
  \subfigure[$\tan\beta = 2.0$, $\kappa=1.0$, $\mu = 200~{\rm GeV}$]{\includegraphics[width=9cm]{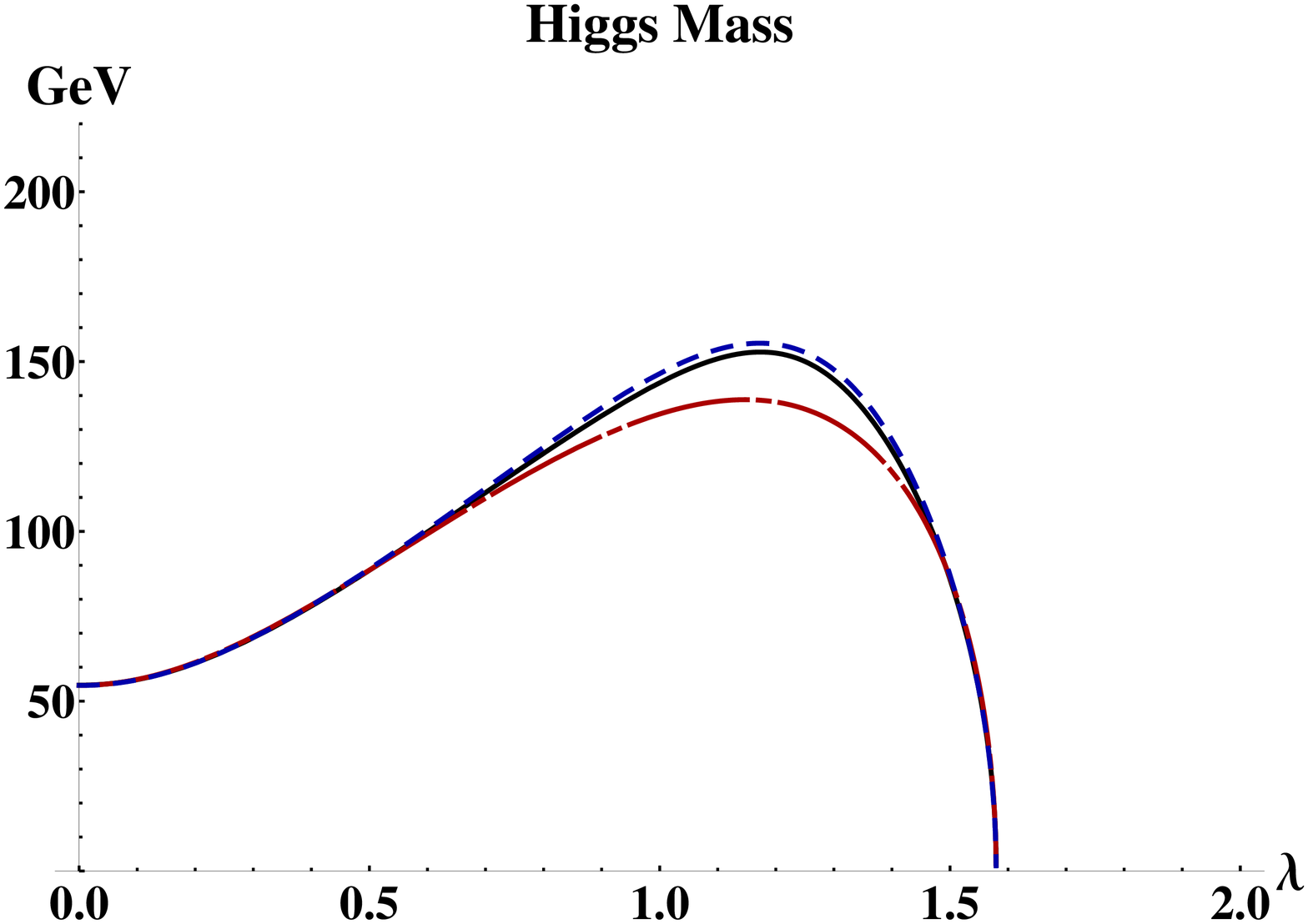}}
\caption{The Higgs boson mass as a function of $\lambda$ for fixed values 
 of $\tan\beta$ and $\kappa$, given by the exact tree-level formula (black, 
 solid line), the first-order (red, dot-dashed) and second-order (blue, 
 dashed) analytical formulae in Eqs.~(\ref{eq:first},~\ref{eq:second}). 
 On the left, the piece-wise exact analytical solutions for $\tan\beta 
 = 1$ are also shown as magenta, dotted lines.  The second-order formula 
 gives a very good fit away from the point $\tan\beta =1$, $\lambda = 
 \lambda_{\rm crit}$ where the dotted lines cross.}
\label{fig:analyt}
\end{figure}
We first describe the effects of $\lambda$-doublet mixing in the 
non-decoupling regime, $2 \lambda^2 v_u v_d \sim B_\mu + v_u v_d 
\bar{g}^2/2$, in terms of an expansion in $m_h^2/m_H^2$ (light Higgs 
mass over heavy Higgs mass) up to second order.  For this purpose, we 
suspend doublet-singlet mixing in this paragraph; it will be discussed 
below.  We find that to first order in the above mentioned expansion, 
the light Higgs mass is given by
\begin{equation}
  m_h^2 \approx M_Z^2 \cos^2\! 2\beta + \lambda^2 v^2 \sin^2\! 2\beta 
    \left( 1 - \frac{\lambda^2 v^2 - M_Z^2}{2 B_\mu} \sin 2\beta \right),
\label{eq:first}
\end{equation}
where we have used the zeroth order result $m_H^2 = 2 B_\mu/\sin 2\beta$. 
This approximation is valid to within $10\%$ for $\tan\beta \simgt 2$ 
with $\kappa = 1$, $\mu = 200~{\rm GeV}$.  Performing the expansion to 
second order in $m_h^2/m_H^2$, we obtain
\begin{equation}
  m_h^2 \approx m_{h,0}^2\, 
    \frac{2B_\mu - M_Z^2 \sin 2\beta}{2B_\mu - m_{h,0}^2 \sin 2\beta},
\label{eq:second}
\end{equation}
where $m_{h,0}^2$ is given by Eq.~(\ref{eq:first}).  We find that this 
second-order expansion gives the correct Higgs mass to within $2\%$, 
$5\%$, and $10\%$ for $\tan\beta > 2$, $1.4$, and $1.2$, respectively, 
with $\kappa = 1$, $\mu = 200~{\rm GeV}$.  The analytical formulae are 
compared with the exact tree-level values in Fig.~\ref{fig:analyt}. 
As $\tan\beta$ approaches one, the gap between the light and heavy Higgs 
masses shrinks to zero at a value of $\lambda = \lambda_{\rm crit}$ 
given by $\lambda_{\rm crit}^2 v^2 = B_\mu + M_Z^2/2$.  This kink-structure 
cannot be faithfully described by a perturbative expansion in $m_h^2/m_H^2$. 
For $\tan\beta = 1$, Eq.~(\ref{eq:decoup}) is an exact solution 
for $\lambda < \lambda_{\rm crit}$, while $m_h^2 = 2B_\mu + M_Z^2 
- \lambda^2 v^2$ for $\lambda > \lambda_{\rm crit}$.  In the case 
of a large Higgs mass, $m_h^2 \gg M_Z^2$, a useful expression is
\begin{equation}
  m_h^2 = \frac{1}{2} \left( \frac{B_\mu}{\sin\beta \cos\beta} 
    - \sqrt{\left(2 \lambda^2 v^2 \sin 2\beta - 2 B_\mu \right)^2 
    + 4 B_\mu^2 \cot^2\! 2\beta} \right).
\label{eq:exact}
\end{equation}

We now give an analytic formula for the correction to the Higgs mass 
from mixing with the singlet in the limit of negligible $\lambda$-doublet 
mixing, $2 \lambda^2 v_u v_d \ll B_\mu + \bar{g}^2 v_u v_d/2$, which 
corresponds to the doublet-doublet decoupling regime.  Performing 
again an expansion in $m_h^2/m_s^2$, with $m_s^2$ the ${\cal M}^2_{33}$ 
entry of the scalar mass matrix, one finds to first order
\begin{equation}
  \delta m_h^2 = -\frac{4 \mu^2 v^2}{m_s^2} \left(\lambda - \Bigl( 
    \kappa + \lambda \frac{A_\lambda}{2\mu} \Bigr) \sin 2\beta \right)^2.
\label{eq:sd-mix}
\end{equation}
In the limit $A_{\lambda,\kappa} \rightarrow 0$, this agrees with the 
result in Ref.~\cite{Ellwanger:2009dp}.

\end{document}